\documentclass[3p,times,twocolumn]{elsarticle}
\usepackage{ecrc-ac}
\usepackage{color}
\usepackage[rflt]{floatflt}
\usepackage{float}
\usepackage{amsmath}
\biboptions{sort&compress}

\newcommand{\gsim}{\raisebox{-0.07cm   }{$\, \stackrel{>}{{\scriptstyle\sim}}\, $}}
\newcommand{\N}{\nonumber}
\newcommand{\Li}{{\rm Li}}

\newcommand{\Ahathat}{\hat{\hspace*{-1mm}\hat{A}}}

\newcommand{\Atil}{\tilde{A}}
\newcommand{\Ctil}{\tilde{C}}

\newcommand{\ep}{\varepsilon}

\allowdisplaybreaks[4]

\volume{00}

\firstpage{1}

\journalname{Nuclear Physics B Proceedings Supplement}

\runauth{J. Bl\"umlein et al.}

\jid{nuphbp}

\jnltitlelogo{Nuclear Physics B Proceedings Supplement}

\usepackage{amssymb}
\usepackage{amsthm}
\usepackage{amsmath}
\newcommand{\Mvec}{{\bf \rm M}}
\newcommand{\GeV}{{\rm GeV}}

\usepackage[figuresright]{rotating}

\begin{document}

\begin{frontmatter}



\dochead{}

\title{{\footnotesize DESY 14-159, DO--TH 14/22, SFB/CPP--14--73 , LPN 14--113}\\
Higher Order Heavy Quark Corrections to Deep-Inelastic Scattering}


\author{J. Bl\"umlein$^a$, A.~De~Freitas$^{a,b}$, and C.~Schneider$^b$}

\address{$^a$Deutsches Elektronen-Synchrotron, DESY, Platanenallee 6, D-15738 
Zeuthen, Germany}

\address{$^b$Research
Institute for Symbolic Computation (RISC), Johannes Kepler
University, Altenbergerstra\ss{}e 69, A-4040 Linz, Austria}

\begin{abstract}
\noindent
The 3-loop heavy flavor corrections to deep-inelastic scattering are
essential for consistent next-to-next-to-leading order QCD analyses.
We report on the present status of the calculation of these corrections
at large virtualities $Q^2$. We also describe a series of mathematical,
computer-algebraic and combinatorial methods  and special function spaces, 
needed to perform these calculations. Finally, we briefly discuss the status 
of measuring $\alpha_s(M_Z)$, the charm quark mass $m_c$, and the parton 
distribution functions at next-to-next-to-leading order from the world 
precision data on deep-inelastic scattering.
\end{abstract}


\end{frontmatter}


\section{Introduction}
\label{sec:1}

\noindent
Deep-inelastic scattering (DIS) provides an important method to determine the distribution function of the partons inside nucleons
and to measure the strong coupling constant $\alpha_s$ from the scaling violations of the nucleon structure functions. Their
heavy flavor corrections also allow the determination of the charm quark mass $m_c$. Currently the world data on DIS have reached a 
precision which needs the next-to-next-to-leading order (NNLO) QCD corrections for an adequate description of the data. While the
massless 3-loop corrections are known \cite{WILS2,Moch:2004pa,Vogt:2004mw,Vermaseren:2005qc}, the heavy flavor Wilson coefficients 
are only known to next-to-leading order \cite{Laenen:1992zk}\footnote{For a numerical implementation in Mellin space see 
\cite{Alekhin:2003ev}.}. Currently the NNLO corrections to the heavy flavor Wilson coefficients are being calculated. In this 
survey we 
give a detailed account on this project, including technical aspects of the computation, such as systematic analytic summation and 
integration techniques, and their computer-algebraic implementation. On very general grounds, Feynman integrals may be turned into nested
sums \cite{Blumlein:2009ta,Blumlein:2010zv}. In this context also new mathematical function spaces and algebras are of importance, 
which have been worked out along with the current project. We also give a brief account on these structures\footnote{For recent reviews
see \cite{Ablinger:2013jta,Ablinger:2013eba}.}. Finally, we present recent phenomenological applications determining the twist-2 parton 
distribution functions at NNLO in the unpolarized case and at NLO in the polarized case and measuring the strong coupling constant
$\alpha_s(M_Z^2)$ and the charm quark mass $m_c$ from the world deep-inelastic scattering data. This survey summarizes main results 
of the research performed in project B3:`Structure functions in the continuum' of  DFG Sonderforschungsbereich Transregio 9, 
Computergest\"utzte Theoretische Teilchenphysik. Related studies using lattice gauge methods are given in \cite{JANSEN}.

The deep-inelastic scattering cross section is of the form
\begin{eqnarray}
\frac{d^2 \sigma}{dx dQ^2} \sim L^{\mu\nu} W_{\mu\nu},
\end{eqnarray}
with $x$ the momentum fraction of the parton in the nucleon, $q^2 = -Q^2$ the virtuality of the exchanged gauge boson, and $L^{\mu\nu}$ and
$W_{\mu\nu}$ the leptonic and hadronic tensors. While the leptonic tensor is a perturbative quantity, the hadronic tensor is not.
In the unpolarized electro-magnetic case it is parameterized as \cite{Blumlein:2012bf}

\vspace*{-3mm}
{\small
\begin{eqnarray}
\label{eq:HAD}
W_{\mu\nu} &=& \frac{1}{4\pi} \int d^4 \xi \exp(i q \xi) \langle P,s| [J^{\rm em}_\mu(\xi), J^{\rm em}_\nu(0)|P,s\rangle
\nonumber\\ 
&=& \frac{1}{2x} \left(g_{\mu\nu} - \frac{q_\mu q_\nu}{q^2}\right) F_L(x,Q^2) 
+ \frac{2x}{Q^2} \Bigl(P_\mu P_\nu  + 
\nonumber\\ &&
\frac{q_\mu P_\nu + 
q_\nu P_\mu}{2x} - \frac{Q^2}{4 
x^2} g_{\mu\nu} \Bigr) F_2(x,Q^2).
\end{eqnarray}
}

\noindent
The structure functions $F_{2,L}(x,Q^2)$ in (\ref{eq:HAD}) are non-perturbative inclusive quantities and contain light and 
heavy flavor contributions. At sufficiently high values of $Q^2$, where the leading twist approximation is valid, the structure 
functions can be represented by 
{\small
\begin{eqnarray}
F_{2,L}(x,Q^2) = \sum_j \mathbb{C}_{(2,L),j}\left(x,\frac{Q^2}{\mu^2}, \frac{m_i^2}{Q^2}\right) \otimes f_j(x,\mu^2)
\end{eqnarray}
}

\noindent
in the $\overline{\rm MS}$-scheme.\footnote{For genuine scheme-invariant representations of structure functions see 
e.g.~\cite{Blumlein:2000wh}.}
Here $\mathbb{C}_j$ denote the Wilson coefficients, $\mu$ the factorization scale, $m_i$ the heavy quark masses, $f_j$ are the twist-2
parton densities and $\otimes$ is the Mellin convolution
{\small
\begin{eqnarray}
\hspace*{-6mm}
\left[A \otimes B\right](x) = 
\int_0^1 dx_1 \int_0^1 dx_2 \delta(x - x_1 x_2) A(x_1) B(x_2),
\end{eqnarray}
}

\noindent
which decouples into a product under the Mellin transform

\vspace*{-5mm}
{\small
\begin{eqnarray}
\Mvec[A(x)](N) = \int_0^1 dx x^{N-1} A(x)~.
\end{eqnarray}
}

\noindent
The Wilson coefficients $\mathbb{C}_j$ contain both the contributions due to inclusive massless partons, $C_j$, as well as the  massive quark 
contributions, $H_j$,
{\small
\begin{eqnarray}
\mathbb{C}_j = C_j + H_j~.
\end{eqnarray}
}

\vspace*{-5mm}
\noindent
As has been shown in \cite{Buza:1995ie}, at large scales $Q^2 \gg m^2$, with $m$ the heavy quark mass, the heavy quark contributions
$H_j$ factorize into massive operator matrix elements (OMEs) and the massless Wilson coefficients. In the case of the structure 
function
$F_2(x,Q^2)$ this factorization applies starting from $Q^2/m^2 \gsim 10$, while for the structure function $F_L(x,Q^2)$ 
\cite{Blumlein:2006mh,Behring:2014eya} it applies at much larger scales of $Q^2/m^2 \gsim 800$ only \cite{Buza:1995ie}.

In the following we review the status of the calculation of the 3-loop QCD corrections of the heavy flavor contributions to the
deep-inelastic structure functions at large values of $Q^2$ and of the 3-loop matching coefficients in the variable flavor number scheme (VFNS).
First we give a survey on the basic relations and discuss the calculation of finite Mellin moments \cite{Bierenbaum:2009mv}. We turn then to
the calculation of these quantities for general values of the Mellin variable $N$, which requires techniques very different from those applied 
for the individual moments. We first line out different computational methods and discuss then the analytical results having been obtained so far,
including numerical predictions. The solution of the corresponding Feynman integrals requests special function spaces which we 
review together
with main relations for these quantities. Finally we give a brief summary on the status of the parton distribution functions (PDFs) both in the 
unpolarized and polarized case and on the determination of the strong coupling constant $\alpha_s(M_Z)$ and the charm quark mass $m_c$ using
the world DIS and related hard scattering data.
  
\vspace*{-3mm}
\section{$O(a_s^3)$ Heavy Flavor Corrections to Deep-Inelastic Scattering}
\label{sec:2}

\noindent
The leading order QCD heavy flavor corrections to deep inelastic scattering were calculated in Refs.~\cite{LO} in the later 1970s.
They are given by the leading order gauge boson gluon fusion process. In 1992 the next-to-leading order (NLO) corrections were 
accomplished in 
Refs.~\cite{Laenen:1992zk} in semi-analytic form  and presented for the flavor-tagged process in $x$-space. A numerical 
implementation
of these corrections in Mellin-$N$ has been given in \cite{Alekhin:2003ev}. In the case of the contributions to the inclusive
structure functions $F_{2,L}(x,Q^2)$ further virtual corrections contribute, cf.~\cite{Bierenbaum:2009zt}.

Fully analytic representations at 2-loop order could be obtained in the region of large virtualities using the factorization
\cite{Buza:1995ie}. Results were given in the unpolarized \cite{Buza:1995ie,Buza:1996wv,Bierenbaum:2007dm,Bierenbaum:2007qe,
Bierenbaum:2009zt} and polarized case 
\cite{Buza:1996xr,Bierenbaum:2007pn} for neutral current
interactions and in the charged current case \cite{Buza:1997mg}. The latter results had to be corrected for  part of the terms in 
\cite{Blumlein:2014fqa}.\footnote{For the corresponding 1-loop results, see \cite{Gluck:1996ve,Blumlein:2011zu}.} The calculation
in this case requests proper current crossing, cf.~\cite{CURR}.
As a preparatory step to calculate the 3-loop corrections, the 2-loop OMEs had to
be calculated in the dimensional parameter $\varepsilon = D-4$ up to $O(\varepsilon)$ \cite{Bierenbaum:2008yu}, 
which required more advanced analytic summation technologies.

Let us now turn to the asymptotic heavy flavor Wilson coefficients in the inclusive case and to the OMEs at 3-loop order.

\subsection{General Formalism}

\noindent
The Wilson coefficients in the region $Q^2 \gg m^2$ are given by \cite{Bierenbaum:2009mv}
{\footnotesize
\begin{eqnarray}
\label{eqWIL1}
\hspace*{-7mm}
L_{2,q}^{\sf NS}(N_F) &=&
a_s^2 \left[A_{qq,Q}^{{\sf NS}, (2)}(N_F) +
\hat{C}^{{\sf NS}, (2)}_{2,q}(N_F)\right]
+ a_s^3 \left[A_{qq,Q}^{{\sf NS}, (3)}(N_F) \right.
\nonumber\\
&+& \left.
  A_{qq,Q}^{{\sf NS}, (2)}(N_F) C_{2,q}^{{\sf NS}, (1)}(N_F)
+ \hat{C}^{{\sf NS}, (3)}_{2,q}(N_F)\right]
\\
\label{eqWIL2}
\hspace*{-7mm}
{\tilde{L}}_{2,q}^{\sf PS}(N_F) &=&
a_s^3 \left[~\Atil_{qq,Q}^{{\sf PS}, (3)}(N_F)
+  A_{gq,Q}^{(2)}(N_F) \Ctil_{2,g}^{(1)}(N_F+1) \right.
\nonumber\\ &+& \left.
\hat{\Ctil}^{{\sf PS}, (3)}_{2,q}(N_F)\right]
\\
\label{eqWIL3}
\hspace*{-7mm}
{\tilde{L}}_{2,g}^{\sf S}(N_F) &=&
a_s^2 A_{gg,Q}^{(1)}(N_F) \Ctil_{2,g}^{(1)}(N_F+1)
+ a_s^3 \Bigl[\Atil_{qg,Q}^{(3)}(N_F)
\nonumber\\ &+&
A_{gg,Q}^{(1)}(N_F) \Ctil_{2,g}^{(2)}(N_F+1) 
+  A_{gg,Q}^{(2)}(N_F) \Ctil_{2,g}^{(1)}(N_F+1)
\nonumber\\ && 
+  ~A_{Qg}^{(1)}(N_F) \Ctil_{2,q}^{{\sf PS}, (2)}(N_F+1)
+ \hat{\Ctil}^{(3)}_{2,g}(N_F)\Bigr]
\\
\label{eqWIL4}
\hspace*{-7mm}
H_{2,q}^{\sf PS}(N_F)
&=& a_s^2 \left[A_{Qq}^{{\sf PS}, (2)}(N_F)
+ \Ctil_{2,q}^{{\sf PS}, (2)}(N_F+1)\right]
+ a_s^3 \Bigl[A_{Qq}^{{\sf PS}, (3)}(N_F)
\nonumber\\
&+& 
~\Ctil_{2,q}^{{\sf PS}, (3)}(N_F+1)
+ A_{gq,Q}^{(2)}(N_F) \Ctil_{2,g}^{(1)}(N_F+1)
\nonumber\\ && \hspace*{5mm}
+ A_{Qq}^{{\sf PS},(2)}(N_F) C_{2,q}^{{\sf NS}, (1)}(N_F+1)
\Bigr]
\\
\label{eqWIL5}
\hspace*{-7mm}
H_{2,g}^{\sf S}(N_F) &=& a_s \left[~A_{Qg}^{(1)}(N_F)
+ \Ctil^{(1)}_{2,g}(N_F+1) \right] + a_s^2 \Bigl[A_{Qg}^{(2)}(N_F)
\nonumber\\
&+& 
A_{Qg}^{(1)}(N_F)~C^{{\sf NS}, (1)}_{2,q}(N_F+1)
+ A_{gg,Q}^{(1)}(N_F) \Ctil^{(1)}_{2,g}(N_F+1)
\nonumber\\ &+&
\Ctil^{(2)}_{2,g}(N_F+1) \Bigr]
+ a_s^3 \Bigl[A_{Qg}^{(3)}(N_F)
+A_{Qg}^{(2)}(N_F)
\nonumber\\ && \times
~C^{{\sf NS}, (1)}_{2,q}(N_F+1)
+A_{gg,Q}^{(2)}(N_F) \Ctil^{(1)}_{2,g}(N_F+1)
\nonumber\\ &+&
A_{Qg}^{(1)}(N_F)\left[
C^{{\sf NS}, (2)}_{2,q}(N_F+1)
+ \Ctil^{{\sf PS}, (2)}_{2,q}(N_F+1)\right]
\nonumber\\ &+&
A_{gg,Q}^{(1)}(N_F)~\Ctil^{(2)}_{2,g}(N_F+1)
+\Ctil^{(3)}_{2,g}(N_F+1) \Bigr]~.
\end{eqnarray}
}

\normalsize
\vspace*{-3mm}
\noindent
Here $a_s = \alpha_s/(4 \pi)$ denotes the strong coupling constant and we used the convention {\small $\hat{f}(N_F) = f(N_F+1) - f(N_F), 
\tilde{f}(N_F) = f(N_F)/N_F, \hat{\tilde{f}}(N_F) = 
\widehat{[\tilde{f}]}(N_F)$}. Since the massless Wilson coefficients are known \cite{WILS2,Vermaseren:2005qc}, only the massive
OMEs $A_{ij}$ at 3-loop order have to be calculated. 

In many applications in collider physics one is interested in describing the process of a heavy quark becoming light at large 
virtualities. Here one would like to match the {\it universal} terms for the parton distributions and their
combinations within this transition for one heavy quark turning into an effectively massless state at the time.\footnote{For
the corresponding relations in the case of two heavy quarks becoming light, see Ref.~\cite{BW1}.} The relations between the parton 
distribution
functions for $N_F$ and $N_F+1$ massless quarks, respectively, describing the variable flavor number scheme (VFNS) at a matching 
scale 
$\mu^2$ read \cite{Buza:1996wv,Bierenbaum:2009mv}

\vspace*{-4mm}
{\footnotesize
\begin{eqnarray}
\label{HPDF1}
\lefteqn{ 
\hspace*{-7mm}
\label{eq:VNS}
f_k(N_F+1) + f_{\overline{k}}(N_F+1) =} \nonumber\\ &&
\hspace*{-12mm}
A_{qq,Q}^{\rm NS}\left(N_F,\frac{\mu^2}{m^2}\right)
\cdot \bigl[f_k(N_F) + f_{\overline{k}}(N_F)\bigr]
+ \tilde{A}_{qq,Q}^{\rm
PS}\left(N_F,\frac{\mu^2}{m^2}\right)
\cdot \Sigma(N_F,\mu^2)
\nonumber\\
& & \hspace*{-12mm}
+ \tilde{A}_{qg,Q}\left(N_F,\frac{\mu^2}{m^2}\right)
\cdot G(N_F,\mu^2),
\nonumber\\
\label{fQQB}
\lefteqn{\hspace*{-7mm}
f_Q(N_F) + f_{\overline{Q}}(N_F+1) =} \nonumber\\ &&
\hspace*{-12mm}
{A}_{Qq}^{\rm PS}\left(N_F,\frac{\mu^2}{m^2}\right)
\cdot \Sigma(N_F,\mu^2)
+ {A}_{Qg}\left(N_F,\frac{\mu^2}{m^2},N\right)
\cdot G(N_F,\mu^2),
\nonumber
\end{eqnarray}
\begin{eqnarray}
\lefteqn{\hspace*{-7mm} \Sigma(N_F+1) =} \nonumber\\ &&
\hspace*{-12mm}
\Biggl[A_{qq,Q}^{\rm NS}\left(N_F, \frac{\mu^2}{m^2}\right) +
          N_F \tilde{A}_{qq,Q}^{\rm PS}\left(N_F, \frac{\mu^2}{m^2}\right)
         + {A}_{Qq}^{\rm PS}\left(N_F, \frac{\mu^2}{m^2}\right)
        \Biggr]
\cdot \Sigma(N_F,\mu^2) \nonumber\\
& & \hspace*{-12mm}
+ \left[N_F \tilde{A}_{qg,Q}
\left(N_F,\frac{\mu^2}{m^2}\right) +
          {A}_{Qg}\left(N_F, \frac{\mu^2}{m^2}\right)
\right]
\cdot G(N_F,\mu^2),
\end{eqnarray}
\begin{eqnarray}
\label{HPDF2}
\lefteqn{\hspace*{-7mm} G(N_F+1) =} \nonumber\\   
&&
\hspace*{-12mm}
A_{gq,Q}\left(N_F,
\frac{\mu^2}{m^2}\right)
                    \cdot \Sigma(N_F,\mu^2)
+ A_{gg,Q}\left(N_F, \frac{\mu^2}{m^2}\right)
                    \cdot G(N_F,\mu^2),
\\
\lefteqn{\hspace*{-7mm} \Delta(N_F+1) =} \nonumber\\ && 
\hspace*{-12mm}
  f_k(N_F+1,\mu^2) 
+ f_{\overline{k}}(N_F+1,\mu^2)
- \frac{1}{N_F+1}
\Sigma(N_F+1,\mu^2),
\end{eqnarray}
}

\normalsize

\vspace*{-5mm}
\noindent
with $k = u,d,s$ denoting the massless quark flavors.
We would like to note that usually the appropriate scale $\mu^2$ is different from that of the heavy quark mass $m^2$ 
\cite{Blumlein:1998sh}.

At the 3-loop  level the eight OMEs to be calculated consist of 2964 Feynman diagrams  with local operator insertions \cite{LCE}, cf.
\cite{Bierenbaum:2009mv}. They are generated using an extension \cite{Bierenbaum:2009mv} of {\tt QGRAF} \cite{Nogueira:1991ex}.
The results for the individual diagrams are then written in {\tt FORM} \cite{Tentyukov:2007mu}. The color algebra is carried out
using the code {\tt Color} \cite{vanRitbergen:1998pn}. In the {\tt FORM}-code  \cite{Bierenbaum:2009mv}  the Dirac-algebra is 
performed. It has exits both to the calculation of individual moments and for a structure allowing to calculate them for 
general values of the Mellin variable $N$ in $D$ dimensions. In this way one first obtains the unrenormalized massive OMEs.
In the following we will focus on the case of a single heavy quark mass. Recently, also the case of two heavy quark masses
has been dealt with in \cite{BW1}. Let us illustrate the structure of the unrenormalized OME for the flavor pure singlet case
\cite{Bierenbaum:2009mv,Ablinger:2014nga} as an example~:

\vspace*{-2mm}
{\footnotesize
\begin{eqnarray}
\lefteqn{\hspace*{-5mm} \Ahathat_{Qq}^{(3),{\sf PS}} =} \nonumber\\ &&
\hspace*{-12mm}
     \left(\frac{\hat{m}^2}{\mu^2}\right)^{3\ep/2}\Biggl[
     \frac{\hat{\gamma}_{qg}^{(0)}\gamma_{gq}^{(0)}}{6\ep^3}
                  \Biggl(
                         \gamma_{gg}^{(0)}
                        -\gamma_{qq}^{(0)}
                        +6\beta_0
                        +16\beta_{0,Q}
                  \Biggr)
   +\frac{1}{\ep^2}\Biggl(
                        -\frac{4\hat{\gamma}_{qq}^{(1),{\sf PS}}}{3}
\nonumber\\ &&  \hspace*{-12mm}                                
\times \Bigl[
                                        \beta_0
                                       +\beta_{0,Q}
                                 \Bigr]
                        -\frac{\gamma_{gq}^{(0)}\hat{\gamma}_{qg}^{(1)}}{3}
                        +\frac{\hat{\gamma}_{qg}^{(0)}}{6}
                                 \Bigl[
                                        2\hat{\gamma}_{gq}^{(1)}
                                       -\gamma_{gq}^{(1)}
                                 \Bigr]
                        +\delta m_1^{(-1)} \hat{\gamma}_{qg}^{(0)}
                                           \gamma_{gq}^{(0)}
                 \Biggr)
\nonumber\\ && \hspace*{-12mm}
   +\frac{1}{\ep}\Biggl(
                           \frac{\hat{\gamma}_{qq}^{(2),{\sf PS}}}{3}
                        -N_F\frac{\hat{\tilde{\gamma}}_{qq}^{(2),{\sf PS}}}{3}
                          +\hat{\gamma}_{qg}^{(0)}a_{gq,Q}^{(2)}
                          -\gamma_{gq}^{(0)}a_{Qg}^{(2)}
                          -4(\beta_0+\beta_{0,Q})a_{Qq}^{(2),{\sf PS}}
\nonumber\\ && \hspace*{-12mm}
                   -\frac{\hat{\gamma}_{qg}^{(0)}\gamma_{gq}^{(0)}\zeta_2}{16}
                            \Bigl[
                               \gamma_{gg}^{(0)}
                              -\gamma_{qq}^{(0)}
                              +6\beta_0
                            \Bigr]
                   +\delta m_1^{(0)} \hat{\gamma}_{qg}^{(0)}
                                           \gamma_{gq}^{(0)}
                   -\delta m_1^{(-1)} \hat{\gamma}_{qq}^{(1),{\sf PS}}
                 \Biggr)
\nonumber\\ && \hspace*{-12mm}
   +a_{Qq}^{(3),{\sf PS}}
                              \Biggr]~. \label{AhhhQq3PS} 
\end{eqnarray}
}

\normalsize
\noindent
Here $\gamma_{ij}^{(k)}$ denote anomalous dimensions, $\beta_k$ and $\beta_{k,Q}$ are the expansion coefficients in the massless 
and massive case, $\delta m_i^{(j)}$ are the expansion coefficients of the running mass, and $a_{ij}^{(k)} (\bar{a}_{ij}^{(k)})$ 
are the constant and $O(\varepsilon)$ parts of the lower order unrenormalized OMEs, respectively. From the single pole term 
$O(\varepsilon^{-1})$
the contributions $\propto N_F$ of the 3-loop anomalous dimensions are obtained as a by-product of the calculation. In the case of 
$\gamma_{qq}^{\rm 
PS}$ and $\gamma_{qg}$ the complete anomalous dimensions are obtained.

The renormalization of the massive OMEs comprises four steps \cite{Bierenbaum:2009mv}:
\begin{itemize}
\item mass renormalization \cite{MASS}
\item coupling constant renormalization
\item renormalization of the local operator
\item treatment of the collinear singularities.
\end{itemize}

\noindent
The external legs of the OMEs are massless on-shell partons\footnote{External massive on-shell partons have been considered
in Ref.~\cite{Blumlein:2011mi}.}
and the scale of the diagram is set by the heavy quark mass.
As mass effects are involved, we treat the coupling renormalization using the background field method first \cite{BGF}, which
leads to a MOM scheme \cite{Bierenbaum:2009mv}. The transition to the $\overline{\rm MS}$ scheme is performed using the relation

\vspace*{-3mm}
{\footnotesize
\begin{eqnarray}
\hspace*{-5mm}
Z_g^{\overline{\rm MS}^2}(N_F+1) a_s^{\overline{\rm MS}}(\mu^2)
= Z_g^{\rm MOM^2}(N_F+1) a_s^{\rm MOM}(\mu^2)~.
\end{eqnarray}}

\normalsize
\vspace*{-3mm}
\noindent
The $Z$-factors for the ultraviolet and collinear singularities, unlike in the massless case, are here not inverse to each other. 

\subsection{Moments}

\noindent
We first discuss the calculation of finite moments of the different OMEs. They are calculated projecting the results of the 
{\tt FORM}-calculation on a massive tadpole, which is then evaluated using the code {\tt MATAD} \cite{Steinhauser:2000ry}.
The calculation requests about a factor of 5 more memory and computational time going from $N \rightarrow N+2$. The present 
resources
allow to compute moments up to $N = 10, ..., 14$ depending on the OME \cite{Bierenbaum:2009mv}. 
For transversity \cite{Barone:2001sp}
corresponding results have been derived in \cite{Blumlein:2009rg}.
As a representative of the moments we show the 10th moment of the constant part of the unrenormalized OME $A_{Qg}^{(3)}$, 
$a_{Qg}^{(3)}(N=10)$, \cite{Bierenbaum:2009mv}:

{\footnotesize
\begin{eqnarray}
\lefteqn{\hspace*{-0.7cm}
 a_{Qg}^{(3)}(10) =} \nonumber\\  &&
\hspace*{-1.2cm}
T_FC_A^2
      \Biggl(
                 \frac{6830363463566924692253659}{685850575063965696000000}
                -\frac{563692}{81675}{B_4}
                +\frac{483988}{9075}\zeta_4
\N \\ \N \\ &&
\hspace*{-1.2cm}
                -\frac{103652031822049723}{415451499724800}\zeta_3
                -\frac{20114890664357}{581101290000}\zeta_2
      \Biggr)
\N \\ \N \\ &&
\hspace*{-1.2cm}
+T_FC_FC_A
      \Biggl(
                 \frac{872201479486471797889957487}{2992802509370032128000000}
                +\frac{1286792}{81675}{B_4}
\N \\ \N \\ &&
\hspace*{-1.2cm}
                -\frac{643396}{9075}\zeta_4
                -\frac{761897167477437907}{33236119977984000}\zeta_3
                +\frac{15455008277}{660342375}\zeta_2
      \Biggr)
\N \\ \N \\ &&
\hspace*{-1.2cm}
+T_FC_F^2
      \Biggl(
                -\frac{247930147349635960148869654541}{148143724213816590336000000}
                -\frac{11808}{3025}{B_4}
\N \\ \N \\ &&
\hspace*{-1.2cm}
                +\frac{53136}{3025}\zeta_4
                +\frac{9636017147214304991}{7122025709568000}\zeta_3
                +\frac{14699237127551}{15689734830000}\zeta_2
      \Biggr)
\N \\ \N \\ &&
\hspace*{-1.2cm}
+T_F^2C_A
      \Biggl(
                 \frac{23231189758106199645229}{633397356480430080000}
                +\frac{123553074914173}{5755172290560}\zeta_3
\N \\ \N \\ &&
\hspace*{-1.2cm}
                +\frac{4206955789}{377338500}\zeta_2
      \Biggr)
\N \\ \N \\
&&
\hspace*{-1.2cm}
+T_F^2C_F
      \Biggl(
                -\frac{18319931182630444611912149}{1410892611560158003200000}
                -\frac{502987059528463}{113048027136000}\zeta_3
\N \\ \N\\ 
&&
\hspace*{-1.2cm}
                +\frac{24683221051}{46695639375}\zeta_2
      \Biggr)
                -\frac{896}{1485}T_F^3\zeta_3
\N \\ \N \\ 
&&
\hspace*{-1.2cm}
+N_FT_F^2C_A
      \Biggl(
                 \frac{297277185134077151}{15532837481700000}
                -\frac{1505896}{245025}\zeta_3
                +\frac{189965849}{188669250}\zeta_2
      \Biggr)
\N \\ \N \\ &&
\hspace*{-1.2cm}
+N_FT_F^2C_F
      \Biggl(
                -\frac{1178560772273339822317}{107642563748181000000}
                +\frac{62292104}{13476375}\zeta_3
\N \\ \N \\
&&
\hspace*{-1.2cm}
                -\frac{49652772817}{93391278750}\zeta_2
      \Biggr)
~.
\end{eqnarray}
}

\normalsize
\noindent 
Here $N_F$ denotes the number of massless flavors  and the color factors are $C_A = N_c, C_F = (N_c^2-1)/(2 N_c), T_F = 1/2$ for $SU(N_c)$ and $N_c 
= 3$ for QCD, $\zeta_k = 
\sum_{l=1}^\infty 1/l^k, k 
\geq 2$ denote the values of the Riemann $\zeta$-function at the integers $k$, cf.~\cite{Blumlein:2009cf}, and the constant ${B_4}$ is given 
by
{\small
\begin{eqnarray}
\label{eqB4}
{B_4} = 
-4\zeta_2\ln^2(2) +\frac{2}{3}\ln^4(2)
-\frac{13}{2}\zeta_4
                  +16 {\Li}_4\left(\frac{1}{2}\right),
\end{eqnarray}
}

\normalsize
\noindent
and $\Li_n(x) = \sum_{k=1}^\infty x^k/k^n$ denotes the polylogarithm \cite{LEWIN:1958,LEWIN:1981}.
\subsection{Results for general values of $N$}
\label{sec:2.1}

\noindent
The renormalized OMEs (\ref{AhhhQq3PS}) at 3-loop order have the form

{\small
\begin{eqnarray}
\hspace*{-5mm}
A_{ij}^{(3)} =  \sum_{k=0}^3 \hat{a}_{ij}^{(3,3-k)} \ln^k\left(\frac{m^2}{\mu^2}\right)~,
\end{eqnarray}}

\normalsize
\noindent
with $\mu^2$ the renormalization scale. All logarithmic contributions can be predicted from the 2-loop results and the 3-loop 
anomalous dimensions, as well as a series of contributions to the constant term.  They have been discussed in detail in 
Ref.~\cite{Behring:2014eya}. It remains to calculate the constant term of the unrenormalized OMEs. Before we come to the
physical results we outline the calculation methods used.

\subsubsection{Calculation Methods}
\label{sec:2.1a}

\noindent
In the case of general values of $N$ we have reduced all Feynman diagrams to master integrals by the integration by parts (IBP) 
relations \cite{IBP,Laporta:2001dd} using an extended version of {\tt Reduze2} \cite{vonManteuffel:2012np}\footnote{{\tt Reduze2} 
uses the packages {\tt Fermat} \cite{FERMAT} and {\tt Ginac} \cite{Bauer:2000cp}. IBP-reductions have also been implemented in the 
codes \cite{IBPC}.} allowing for the treatment of local operators, resumming them into a generating function. We generally
refer to this representation in the case of the more involved diagrams, while simpler ones have been calculated directly without 
any 
reduction.

The different Feynman integrals or the master integrals are calculated by the following techniques. In the simpler cases
they integrate to Beta-functions or (generalized) hypergeometric functions \cite{GHYP,Slater}, after suitable binomial
decompositions and using special variable transformations \cite{Hamberg}
{\small
\begin{eqnarray}
&& \hspace*{-1.3cm}
{_2F_1}(a,b;c;x) = \frac{\Gamma(c)}{\Gamma(b) \Gamma(c-b)} \int_0^1 dz z^{b-1} (1-z)^{c-b-1} (1-zx)^{-a}
\nonumber\\
\\
&& \hspace*{-1.3cm}
{_{p+1}F_{q+1}}(c,(a);d,(b);x) = \frac{\Gamma(d)}{\Gamma(d) \Gamma(d-c)} \int_0^1 dz  z^{c-1} 
\nonumber\\ && \hspace*{20mm}
\times (1-z)^{d-c-1} {_pF_q}((a);(b);zx)~.
\end{eqnarray}
}

\noindent
More complicated integrals can be mapped to Appell and Kamp\'{e} de F\'{e}riet functions \cite{Appell}. These functions are 
represented in terms of multiple sums. Subsequently the expansion in the dimensional parameter $\varepsilon = D-4$ has to be 
performed. There are special algorithms to perform the $\varepsilon$-expansion under certain conditions 
\cite{Huber:2005yg,Weinzierl:2002hv,Weinzierl:2004bn,Moch:2005uc,Kalmykov:2007dk,Bytev:2011ks}. However, for general 
summand structures the expansion has been built into the package {\tt Sigma} \cite{SIG1,SIG2}. The multiply nested infinite and finite sums have 
to be evaluated. This is done using the packages {\tt Sigma} based on difference field and ring theory
\cite{Karr:81,Schneider:01,Schneider:05a,Schneider:07d,Schneider:08c,Schneider:10a,Schneider:10b,Schneider:10c,Schneider:13b}, 
{\tt EvaluateMultiSum}, 
{\tt SumProduction} \cite{EMSSP} and {\tt $\rho$Sum} \cite{RHOSUM}. The infinite sums need limiting processes and thus require the asymptotic
expansion of the respective sums. Furthermore, one would like to map to basis representations. This can be done with the help 
of the package {\tt HarmonicSums} \cite{Ablinger:2010kw,Ablinger:2011te,Ablinger:2013cf,Ablinger:2013hcp}.

In general it turns out that the higher transcendental functions which would be needed in representing the occurring Feynman diagrams 
at intermediary steps are not found in the literature, but may be represented in terms of a few Mellin-Barnes integrals \cite{MELB}
{\small
\begin{eqnarray}
\frac{1}{(A+B)^\alpha} = \frac{1}{2 \pi i} \int_{c - i\infty}^{c + i\infty} dz A^z B^{-\alpha-z} \frac{\Gamma(-z) 
\Gamma(\alpha+z)}{\Gamma(\alpha)}
\end{eqnarray}
}

\noindent
using algorithms implemented in \cite{Czakon:2005rk,Smirnov:2009up}.
Taking residues one arrives at nested sums, which may again be solved using {\tt Sigma}.

In some of the cases very large numbers of intermediary nested sums over hypergeometric terms emerge, which leads to very large computation
times for {\tt Sigma}, associated with a large memory request. Since {\tt Sigma} solves sum by sum, a behaviour of 
this kind may be related
to the fact that several sums (integrals) have to be solved at once. At the side of step-by-step summation, multi-summation 
algorithms 
\cite{Wegschaider:1997,Blumlein:2010zv}
may be applied. We have successfully computed the corresponding integrals using an implementation of the multi-variate
Almkvist-Zeilberger theorem \cite{AZ,Ablinger:2013hcp}. It allows for multivariate integrals of the type
{\small
\begin{eqnarray}
\hspace*{-6mm}
\prod_{i=1}^k \int_{a_i}^{b_i} d x_i (P(x_n))^N \prod_{l=1}^m (p_l(x_n))^{r_l},~~~N \in \mathbb{N} \backslash \{0\}, r_l \in \mathbb{R}
\end{eqnarray}
}

\noindent
to derive a difference equation in $N$ with polynomial coefficients in $N$ and $\varepsilon$, $Q_k$, 
{\small
\begin{eqnarray}
\label{AZeq}
\hspace*{-6mm}
\sum_{k=0}^l Q_k(N,\varepsilon) F(N+k)  = 0.
\end{eqnarray}
}

\noindent
Here $P(x_i)$ and $p_l(x_i)$ are polynomials and $r_l$ will contain the parameter $\varepsilon$. A term
$(p_l(x_i))^r$ is called hyper-exponential. Sometimes it may be practical to search for inhomogeneous difference
equations instead of (\ref{AZeq}). The corresponding difference equations are solved using {\tt Sigma}, and sequentially
expanding in $\varepsilon$.

A further method consists in deriving systems of linear differential equations for the master integrals \cite{DEQ}.
In part they may be obtained by the IBP-relations having been derived before. These systems may be turned into systems 
of difference equations and solved using the packages {\tt Sigma} and {\tt OreSys} \cite{ORESYS}. We have also used the 
method of hyperlogarithms \cite{Brown:2008um}, extended to the case of massive Feynman diagrams containing local 
operator insertions \cite{Ablinger:2012qm,Ablinger:2014yaa}, by which a series of diagrams having no poles in $\varepsilon$ can be 
calculated. Furthermore, guessing methods \cite{GUESS} are helpful to set up difference equations in a series of cases, see also 
\cite{Blumlein:2009tj}.

In complicated cases several of the above methods are combined. It is needless to say that both the direct calculation of different 
Feynman integrals 
as well as the calculation of the master integrals may require computer memory in the multi-ten Gbyte region and run-times of several 
days to weeks. 
\subsubsection{3-Loop anomalous dimensions}

\noindent
The contributions $\propto N_F$ of the 3-loop anomalous dimensions can be obtained from the single pole terms
of the unrenormalized OMEs. Fixed moments have been computed in \cite{Bierenbaum:2009mv} in the present massive
calculation and found to agree with previous results given in \cite{ANDI3}. For general values of $N$
we show as one example the result in the pure singlet case
 
\vspace*{-5mm}
{\footnotesize
\begin{eqnarray}
\label{eq:gPS3L}
\lefteqn{\hspace*{-7mm} \gamma_{qq}^{(2), \rm PS}(N) = \tfrac{1}{2}[1 + (-1)^N]
\Biggl\{\textcolor{black}{C_F^2 T_F N_F}
\Biggl\{
-\frac{8 \big(N^2+N+2\big) Q_1 S_1^2}
      {(N-1) N^3 (N+1)^3 (N+2)}
} \nonumber\\ &&
\hspace*{-14mm}
+\frac{32 Q_8 S_1}
      {(N-1) N^4 (N+1)^4 (N+2)^3}
-\frac{8 Q_{12}}{(N-1) N^5 (N+1)^5 (N+2)^3}
\nonumber\\ &&
\hspace*{-14mm}
-\frac{8 Q_6 S_2}{(N-1) N^3 (N+1)^3 (N+2)^2} 
+ 32 F
\Biggl[
\frac{1}{3} S_1^3
-S_2 S_1
-\frac{7}{3} S_3
+2 S_{2,1}
\nonumber\\ &&
\hspace*{-14mm}
+ 6\zeta_3
\Biggr]
\Biggr\}
+
\textcolor{black}{C_F T_F^2 N_F^2}
\Biggl\{
-\frac{64 Q_9}{27 (N-1) N^4 (N+1)^4 (N+2)^3}
\nonumber\\ &&
\hspace*{-14mm}
+\frac{64 Q_7 S_1}{9 (N-1) N^3 (N+1)^3 (N+2)^2}
- F \frac{32}{3} [S_1^2 + S_2]
\Biggr\}
\nonumber\\ &&
\hspace*{-14mm}
+
\textcolor{black}{C_F C_A T_F N_F}
\Biggl\{
 \frac{8 \big(N^2+N+2\big) Q_3 S_1^2}{3 (N-1)^2 N^3 (N+1)^3 (N+2)^2} 
\nonumber\\ &&
\hspace*{-14mm}
-\frac{16 Q_{11} S_1}{9 (N-1)^2 N^4 (N+1)^4 (N+2)^2} 
+\frac{16 Q_{13}}{27 (N-1)^2 N^5 (N+1)^5 (N+2)^4}
\nonumber\\ &&
\hspace*{-14mm}
+(-1)^N
\nonumber\\ &&
\hspace*{-14mm}
\times \Biggl[
\frac{128 Q_2 S_1}{3 (N-1) N^2 (N+1)^3 (N+2)^3} 
-\frac{128 Q_{10}}{9 (N-1) N^3 (N+1)^5 (N+2)^4}
\Biggr]
\nonumber\\ &&
\hspace*{-14mm}
+\frac{8 \big(N^2+N+2\big) Q_4 S_2}{3 (N-1)^2 N^3 (N+1)^3 (N+2)^2} 
+\frac{32 Q_5 S_{-2}}{(N-1) N^3 (N+1)^3 (N+2)^2} 
\nonumber\\ &&
\hspace*{-14mm}
+\frac{16 \big(N^2+N+2\big) \big(23 N^2+23 N+58\big)}{3 (N-1) N^2 (N+1)^2 (N+2)} S_3
\nonumber\\ &&
\hspace*{-14mm}
+\frac{32 \big(N^2+N+2\big) \big(7 N^2+7 N+10\big)S_{-3}}{(N-1) N^2 (N+1)^2 (N+2)} 
\nonumber\\ &&
\hspace*{-14mm}
-\frac{64 \big(N^2+N+2\big) \big(3 N^2+3 N+2\big)S_{-2,1}}{(N-1) N^2 (N+1)^2 (N+2)} 
\nonumber\\ &&
\hspace*{-14mm}
+ 32 F
\Biggl[
-\frac{1}{3} S_1^3
+ S_2 S_1
+2 S_{-2} S_1
-2 S_{2,1}
-6\zeta_3
\Biggr]
\Biggr\}
\Biggr\}.
\label{anomPS}
\end{eqnarray}
}

\normalsize
\vspace*{-5mm}
\noindent
Here $F$ and $Q_i$ denote polynomials in $N$, see \cite{Ablinger:2014nga} and $S_{\vec{a}} \equiv S_{\vec{a}}(N)$ are harmonic sums, cf. 
(\ref{eq:HS1}).
Eq.~(\ref{anomPS}) agrees with the result of \cite{Vogt:2004mw} and has been firstly recalculated in \cite{Ablinger:2014nga}.
We also recalculated the contributions $\propto N_F$ to $\gamma_{gq}^{(3)}$ \cite{Ablinger:2014lka}, $\gamma_{qq}^{(3), \rm NS}$ 
and 
for the transversity anomalous dimension $\gamma_{qq}^{(3,\rm NS,TR)}$ \cite{Ablinger:2014vwa}. In the latter case it was the first 
calculation ab initio,
while previously the anomalous dimension had been calculated referring to the principle of maximal transcendentality in 
\cite{Velizhanin:2012nm}. The 2-loop anomalous dimensions \cite{VEC2,TRAN2} are obtained completely.
All results agree with the foregoing literature 
\cite{Moch:2004pa,Vogt:2004mw,ANDI3,Blumlein:2004xt,GRAC,Bagaev:2012bw}.
\subsubsection{The operator matrix element at general values of the Mellin variable $N$}

\noindent
Complete results for general value of $N$ have been obtained for all the leading $N_F$ contributions to the OMEs 
\cite{Ablinger:2010ty,Blumlein:2012vq}, the bubble topologies \cite{Behring:2013dga} and the $O(T_F^2)$ contributions
to the OME $A_{gg}^{(3)}$ \cite{Ablinger:2014uka}. This includes also the asymptotic Wilson coefficients $L_{2,q}^{\rm PS}$ and 
$L_{2,g}^{\rm S}$. While in the former cases all terms can be expressed in terms of nested harmonic sums 
\cite{Vermaseren:1998uu,Blumlein:1998if}, in the contributions of $O(T_F^2)$ \cite{Ablinger:2014uka} also binomially weighted sums 
\cite{Ablinger:2014bra} contribute. The corresponding term is given by
\begin{eqnarray}
\binom{2N}{N} \frac{1}{4^N} \sum_{k=1}^N \left(\frac{4^k S_1(k-1)}{k^2 \binom{2k}{k}} - 7 \zeta_3\right).
\end{eqnarray}

The 3-loop OMEs $A_{qq,Q}^{(3),\rm NS}, A_{qq,Q}^{(3),\rm NS, TR}$ and the Wilson coefficient $L_{2,q}^{\rm NS}$ and the OME
$A_{gq,Q}^{(3)}$ have been calculated in Refs.~\cite{Ablinger:2014vwa,Ablinger:2014lka} and the pure singlet OME 
$A_{Qq}^{(3),\rm PS}$ and Wilson coefficient in $H_{2,q}^{\rm PS}$ in Ref.~\cite{Ablinger:2014nga}. In the former cases
the nested harmonic sums are sufficient to represent the OMEs and asymptotic Wilson coefficients. In the latter case
also generalized harmonic sums \cite{Moch:2001zr,Ablinger:2013cf} contribute. As an example we show the constant part of
the unrenormalized massive pure singlet OME at 3-loop order:

{\footnotesize
\begin{eqnarray}
\label{eq:aQq3PS}
\lefteqn{\hspace*{-7mm} a_{Qq}^{(3), \rm PS}(N) =}
\nonumber \\ &&
\hspace*{-12mm}
\textcolor{black}{C_F T_F^2} \Biggl[
        \frac{32}{27 (N-1) (N+3) (N+4) (N+5)} \biggl(
                \frac{P_{15}}{N^3 (N+1)^2 (N+2)^2} S_2
\nonumber \\ &&
\hspace*{-12mm}
                -\frac{P_{19}}{N^3 (N+1)^3 (N+2)^2} S_1^2
                +\frac{2 P_{28}}{3 N^4 (N+1)^4 (N+2)^3} S_1
\nonumber \\ &&
\hspace*{-12mm}
                -\frac{2 P_{32}}{9 N^5 (N+1)^4 (N+2)^4}
        \biggr)
        -\frac{32 P_3}{9 (N-1) N^3 (N+1)^2 (N+2)^2} \zeta_2
\nonumber \\ &&
\hspace*{-12mm}
        +\biggl(
                \frac{32}{27} S_1^3
                -\frac{160}{9} S_1 S_2
                -\frac{512}{27} S_3
                +\frac{128}{3} S_{2,1}
                +\frac{32}{3} S_1 \zeta_2
                -\frac{1024}{9} \zeta_3
        \biggr) F
\Biggr]
\nonumber \\ &&
\hspace*{-12mm}
+\textcolor{black}{C_F N_F T_F^2} \Biggl[
        \frac{16 P_7}{27 (N-1) N^3 (N+1)^3 (N+2)^2} S_1^2
\nonumber 
\\ 
&&
\hspace*{-12mm}
        +\frac{208 P_7 S_2}{27 (N-1) N^3 (N+1)^3 (N+2)^2} 
        -\frac{32 P_{21} S_1}{81 (N-1) N^4 (N+1)^4 (N+2)^3} 
\nonumber \\ &&
\hspace*{-12mm}
        +\frac{32 P_{29}}{243 (N-1) N^5 (N+1)^5 (N+2)^4}
        +\biggl(
                -\frac{16}{27} S_1^3
                -\frac{208}{9} S_1 S_2
                -\frac{1760}{27} S_3
\nonumber \\ &&
\hspace*{-12mm}
                -\frac{16}{3} S_1 \zeta_2
                +\frac{224}{9} \zeta_3
        \biggr) F
        +\frac{1}{(N-1) N^3 (N+1)^3 (N+2)^2} \frac{16 P_7}{9} \zeta_2
\Biggr]
\nonumber \\ &&
\hspace*{-12mm}
+\textcolor{black}{C_F^2 T_F} \Biggl[
        \frac{32 P_9 S_{2,1}}{3 (N-1) N^3 (N+1)^3 (N+2)^2} 
\nonumber \\ &&
\hspace*{-12mm}
        -\frac{16 P_{14} S_3}{9 (N-1) N^3 (N+1)^3 (N+2)^2} 
        -\frac{4 P_{17} S_1^2}{3 (N-1) N^4 (N+1)^4 (N+2)^3} 
\nonumber \\ &&
\hspace*{-12mm}
        +\frac{4 P_{23} S_2}{3 (N-1) N^4 (N+1)^4 (N+2)^3} 
        +\frac{4 P_{31}}{3 (N-1) N^6 (N+1)^6 (N+2)^4}
\nonumber \\ &&
\hspace*{-12mm}
        +\biggl(
                \biggl(
                        \frac{2 P_5}{N^2 (N+1)^2}
                        -\frac{4 P_1}{N (N+1)} S_1
                \biggr) \zeta_2
                -\frac{4 P_1}{9 N (N+1)} S_1^3
        \biggr) G
\nonumber \\ &&
\hspace*{-12mm}
        +\biggl(
                \biggl(
                        \frac{80}{9} S_3
                        -64 S_{2,1}
                \biggr) S_1
                -\frac{2}{9} S_1^4
                -\frac{20}{3} S_1^2 S_2
                +\frac{46}{3} S_2^2
                +\frac{124}{3} S_4
\nonumber \\ &&
\hspace*{-12mm}
                +\frac{416}{3} S_{2,1,1}
                +64 \biggl(
                        \biggl(
                                S_3({{2}})
                                -S_{1,2}({{2,1}})
                                +S_{2,1}({{2,1}})
\nonumber \\ &&
\hspace*{-12mm}
                                -S_{1,1,1}({{2,1,1}})
                        \biggr) S_1\left({{\frac{1}{2}}}\right)
                        -S_{1,3}\left({{2,\frac{1}{2}}}\right)
                        +S_{2,2}\left({{2,\frac{1}{2}}}\right)
                        -S_{3,1}\left({{2,\frac{1}{2}}}\right)
\nonumber \\ &&
\hspace*{-12mm}
                        +S_{1,1,2}\left({{2,\frac{1}{2},1}}\right)
                        -S_{1,1,2}\left({{2,1,\frac{1}{2}}}\right)
                        -S_{1,2,1}\left({{2,\frac{1}{2},1}}\right)
\nonumber \\ &&
\hspace*{-12mm}
                        +S_{1,2,1}\left({{2,1,\frac{1}{2}}}\right)
                        -S_{2,1,1}\left({{2,\frac{1}{2},1}}\right)
                        -S_{2,1,1}\left({{2,1,\frac{1}{2}}}\right)
\nonumber \\ &&
\hspace*{-12mm}
                        +S_{1,1,1,1}\left({{2,\frac{1}{2},1,1}}\right)
                        +S_{1,1,1,1}\left({{2,1,\frac{1}{2},1}}\right)
                        +S_{1,1,1,1}\left({{2,1,1,\frac{1}{2}}}\right)
                \biggr)
\nonumber \\ &&
\hspace*{-12mm}
                +\biggl(
                        12 S_2
                        -4 S_1^2
                \biggr) \zeta_2
                +\biggl(
                        \frac{112}{3} S_1
                        -448 S_1\left({{\frac{1}{2}}}\right)
                \biggr) \zeta_3
                +144 \zeta_4
                -32 B_4
        \biggr) F
\nonumber \\ &&
\hspace*{-12mm}
        +\frac{32 P_2 2^{-N}}{(N-1) N^3 (N+1)^2} \biggl(
                -S_3({{2}})
                +S_{1,2}({{2,1}})
                -S_{2,1}({{2,1}})
\nonumber \\ &&
\hspace*{-12mm}
                +S_{1,1,1}({{2,1,1}})
                +7 \zeta_3
        \biggr)
        +\biggl(
                -\frac{4 P_8}{3 (N-1) N^3 (N+1)^3 (N+2)^2} S_2
\nonumber \\ &&
\hspace*{-12mm}
                +\frac{8 P_{27}}{3 (N-1) N^5 (N+1)^5 (N+2)^4}
        \biggr) S_1
        +\frac{4 P_{16} \zeta_3}{3(N-1) N^3 (N+1)^3 (N+2)^2} 
\Biggr]
\nonumber \\ &&
\hspace*{-12mm}
+\textcolor{black}{C_A C_F T_F} 
\nonumber \\ &&
\hspace*{-12mm}
\times
\Biggl[
        -\frac{8 P_{10} S_{2,1}}{3 (N-1) N^3 (N+1)^3 (N+2)^2} 
        +\frac{8 P_{12} S_{-3}}{3 (N-1) N^3 (N+1)^3 (N+2)^2} 
\nonumber \\ &&
\hspace*{-12mm}
        +\frac{16 P_{13} S_{-2,1}}{3 (N-1) N^3 (N+1)^3 (N+2)^2} 
        +\frac{8 P_{22} S_3}{27 (N-1)^2 N^3 (N+1)^3 (N+2)^2} 
\nonumber 
\end{eqnarray}
\begin{eqnarray}
&&
\hspace*{-12mm}
        -\frac{4 \left(P_{24} S_1^2 + P_{26} S_2\right)}{27 (N-1)^2 N^4 (N+1)^4 (N+2)^3} 
\nonumber \\ &&
\hspace*{-12mm}
        -\frac{8 P_{33}}{243 (N-1)^2 N^6 (N+1)^6 (N+2)^5}
\nonumber \\ &&
\hspace*{-12mm}
        +\biggl(
                \frac{4 P_4 S_1^3}{27 (N-1) N (N+1) (N+2)} 
                +\biggl(
                        \frac{8}{9} \big(137 N^2+137 N+334\big) S_3
\nonumber \\ &&
\hspace*{-12mm}
                        -\frac{16}{3} \big(35 N^2+35 N+18\big) S_{-2,1}
                \biggr) S_1
                +\frac{8}{3} \big(69 N^2+69 N+94\big) S_{-3} S_1
\nonumber \\ &&
\hspace*{-12mm}
                +\frac{64}{3} \big(7 N^2+7 N+13\big) S_{-2} S_2
                +\frac{2}{3} \big(29 N^2+29 N+74\big) S_2^2
\nonumber \\ &&
\hspace*{-12mm}
                +\frac{4}{3} \big(143 N^2+143 N+310\big) S_4
                -\frac{16}{3} \big(3 N^2+3 N-2\big) S_{-2}^2
\nonumber \\ &&
\hspace*{-12mm}
                +\frac{16}{3} \big(31 N^2+31 N+50\big) S_{-4}
                -8 \big(7 N^2+7 N+26\big) S_{3,1}
\nonumber \\ &&
\hspace*{-12mm}
                -64 \big(3 N^2+3 N+2\big) S_{-2,2}
                -\frac{32}{3} \big(23 N^2+23 N+22\big) S_{-3,1}
\nonumber \\ &&
\hspace*{-12mm}
                +\frac{64}{3} \big(13 N^2+13 N+2\big) S_{-2,1,1}
                +\frac{4 P_4}{3 (N-1) N (N+1) (N+2)} S_1 \zeta_2
\nonumber \\ &&
\hspace*{-12mm}
                -\frac{8}{3} \big(11 N^2+11 N+10\big) S_1 \zeta_3
        \biggr) G
        +\biggl(
                \frac{112}{3} S_{-2} S_1^2
                +\frac{2}{9} S_1^4
                +\frac{68}{3} S_1^2 S_2
\nonumber \\ &&
\hspace*{-12mm}
                -\frac{80}{3} S_{2,1,1}
                +32 \biggl(
                        \biggl(
                                -S_3({{2}})
                                +S_{1,2}({{2,1}})
                                -S_{2,1}({{2,1}})
                                +S_{1,1,1}({{2,1,1}})
                        \biggr) 
\nonumber \\ &&
\hspace*{-12mm}
\times S_1\left({{\frac{1}{2}}}\right)
                        +S_{1,3}\left({{2,\frac{1}{2}}}\right)
                        -S_{2,2}\left({{2,\frac{1}{2}}}\right)
                        +S_{3,1}\left({{2,\frac{1}{2}}}\right)
\nonumber \\ &&
\hspace*{-12mm}
                        -S_{1,1,2}\left({{2,\frac{1}{2},1}}\right)
                        +S_{1,1,2}\left({{2,1,\frac{1}{2}}}\right)
                        +S_{1,2,1}\left({{2,\frac{1}{2},1}}\right)
\nonumber \\ &&
\hspace*{-12mm}
                        -S_{1,2,1}\left({{2,1,\frac{1}{2}}}\right)
                        +S_{2,1,1}\left({{2,\frac{1}{2},1}}\right)
                        +S_{2,1,1}\left({{2,1,\frac{1}{2}}}\right)
\nonumber \\ 
&&
\hspace*{-14mm}
                        -S_{1,1,1,1}\left({{2,\frac{1}{2},1,1}}\right)
                        -S_{1,1,1,1}\left({{2,1,\frac{1}{2},1}}\right)
                        -S_{1,1,1,1}\left({{2,1,1,\frac{1}{2}}}\right)
                \biggr)
\nonumber \\ &&
\hspace*{-14mm}
                +\biggl(
                        4 S_1^2
                        +12 S_2
                        +24 S_{-2}
                \biggr) 
\zeta_2
                +224 S_1\left({{\frac{1}{2}}}\right) \zeta_3
                -144 \zeta_4
                +16 B_4
        \biggl) F
\nonumber \\ &&
\hspace*{-14mm}
        +\frac{16 P_2 2^{-N}}{(N-1) N^3 (N+1)^2} \biggl(
                 S_3({{2}})
                -S_{1,2}({{2,1}})
                +S_{2,1}({{2,1}})
\nonumber \\ &&
\hspace*{-14mm}
                -S_{1,1,1}({{2,1,1}})
                -7 \zeta_3
        \biggr) + S_1 \biggl(
                \frac{4 P_{11} S_2}{9 (N-1)^2 N^3 (N+1)^3 (N+2)^2} 
\nonumber \\ &&
\hspace*{-14mm}
                +\frac{4 P_{30}}{81 (N-1)^2 N^5 (N+1)^5 (N+2)^4}
        \biggr) 
        +
                \frac{32 P_6 S_1 S_{-2}}{3 (N-1) N^3 (N+1)^3 (N+2)^2} 
\nonumber \\ &&
\hspace*{-14mm}
                -\frac{8 P_{18} S_{-2}}{3 (N-1) N^4 (N+1)^4 (N+2)^3}
        -\frac{4 P_{25} \zeta_2}{9 (N-1)^2 N^4 (N+1)^4 (N+2)^3} 
\nonumber \\ &&
\hspace*{-14mm}
        -\frac{8 P_{20} \zeta_3}{9 (N-1)^2 N^3 (N+1)^3 (N+2)^2} 
\Biggr].
\end{eqnarray}}

\normalsize

\vspace*{-3mm}
\noindent
Here $G$ and $P_i$ denote polynomials in $N$ \cite{Ablinger:2014nga}  and $S_{\vec{a}}(\vec{b})$ are generalized harmonic sums, 
cf.~(\ref{eq:GHS1}).
In $x$-space the above contributions due to generalized harmonic sums transform into generalized
harmonic polylogarithms \cite{Ablinger:2013cf}. Their specific combination can be reduced to usual 
harmonic polylogarithms $H_{\vec{a}}$ \cite{Remiddi:1999ew}
of argument $1 - 2x$. In the case of the non-singlet contributions \cite{Ablinger:2014vwa}
the general $N$ result has also been calculated in the polarized case, which is possible solving the $\gamma_5$-problem
via the Ward-Takahashi identity \cite{WT}.

\begin{figure}[H]
\includegraphics[width=0.48\textwidth]{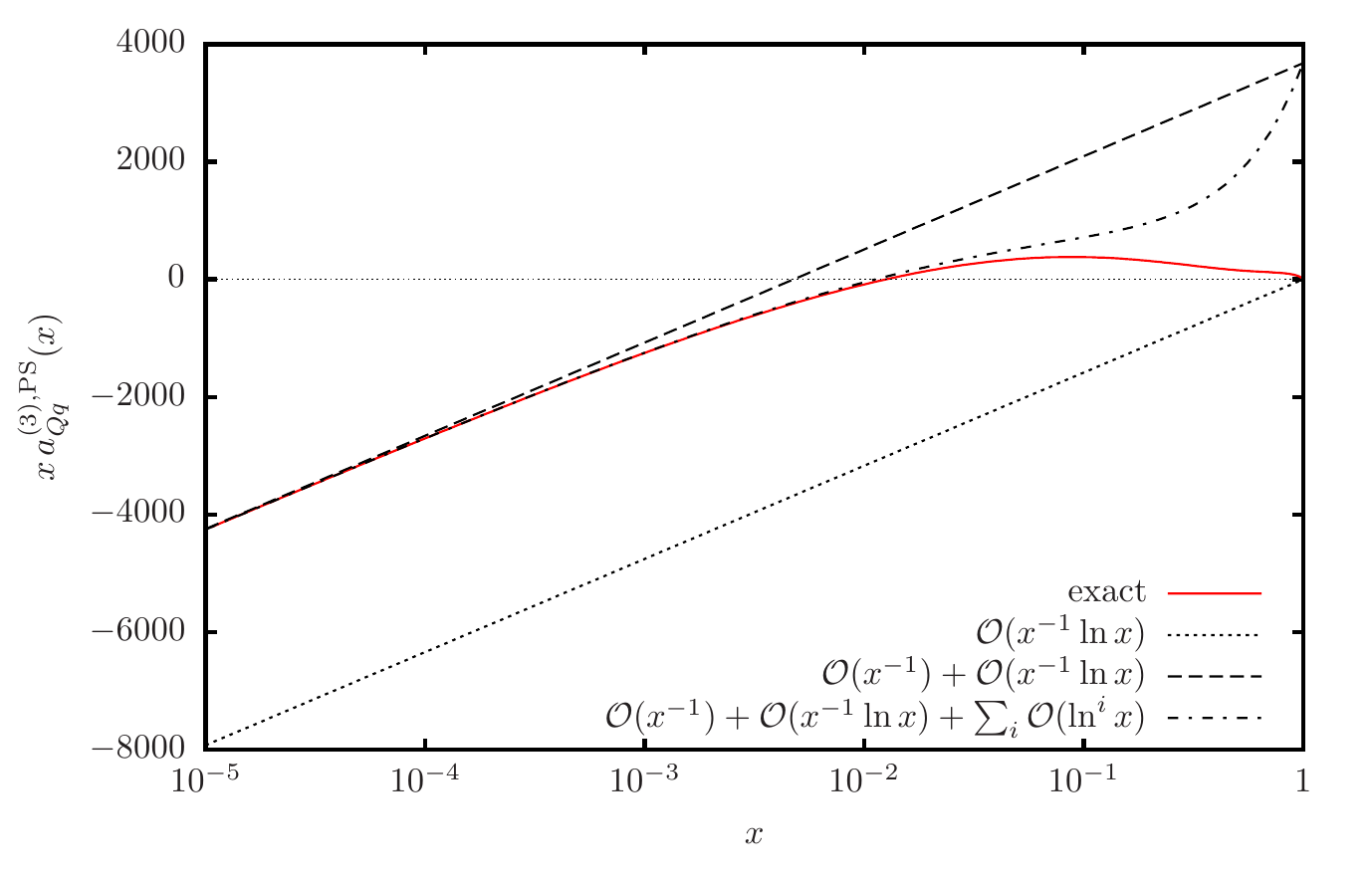}
\caption[]{$xa_{Qq}^{(3),\rm PS}(x)$ in the low $x$ region (solid red line) and leading terms approximating this
quantity; dotted line: `leading' small $x$ approximation $O(\ln(x)/x)$, dashed line: adding the $O(1/x)$-term, dash-dotted
line: adding all other logarithmic contributions; from \cite{Ablinger:2014nga}.}
\label{fig:SX}
\end{figure}

\noindent 
Let us consider the constant part to the unrenormalized OME $A_{Qq}^{\rm PS}$ in $x$-space shown in Figure~\ref{fig:SX}.
The leading small-$x$ contribution can be determined under some assumptions \cite{Kawamura:2012cr} (for a discussion see 
\cite{Ablinger:2014nga}), based on an earlier result in Ref.~\cite{Catani:1990eg}. Our explicit calculation confirms
this leading order prediction. However, it describes the function $a_{Qq}^{(3),\rm PS}$ nowhere.
The next term in the small-$x$
expansion \cite{Ablinger:2014nga} yields a good description at $x \simeq 10^{-4}$ and many more sub-leading terms have to be
added to cover the small-$x$ regime, which is a quite common observation in the case of many quantities \cite{Blumlein:1997em,SMX}.

The calculation of the OMEs $A_{gg,Q}^{(3)}$ and $A_{Qg}^{(3)}$, as well as of the Wilson coefficient $H_{2,g}^{\rm S}$ is currently 
underway. 

\begin{figure}[H]
\includegraphics[width=0.48\textwidth]{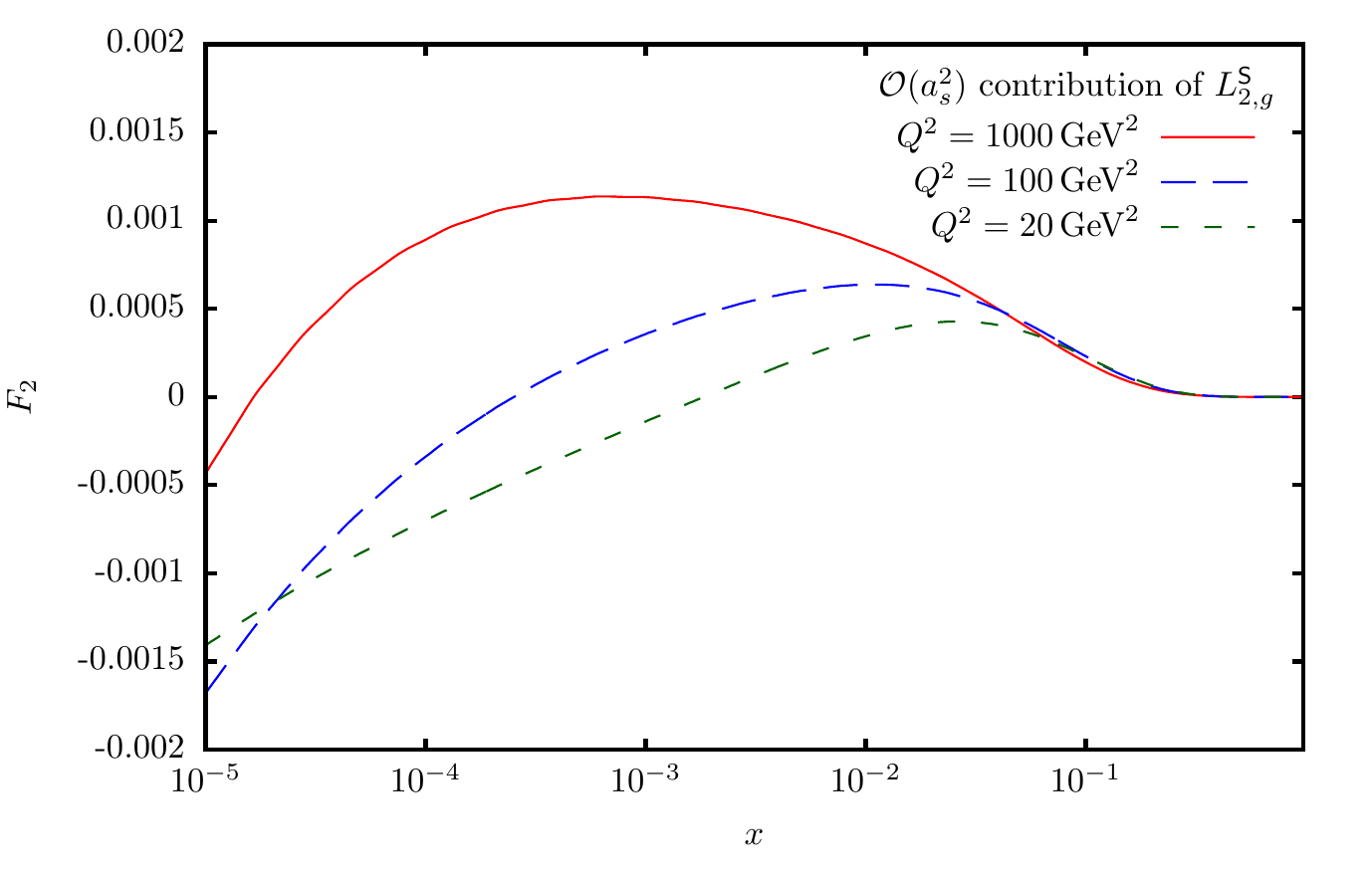}
\caption[]{
The $O(a_s^2)$ contribution by $L_{g,2}^{\sf S}$ to the structure function $F_2(x,Q^2)$ using 
the parton distribution functions \cite{Alekhin:2013nda} and $m_c = 1.59~\GeV$; from \cite{Behring:2014eya}. 
\label{fig1}}
\end{figure}

We now turn to phenomenological applications and consider the contributions of the completed massive Wilson
coefficients in the asymptotic region to the structure function $F_2(x,Q^2)$. In Figure~\ref{fig1} 
the $O(a_s^2)$ contribution by $L_{g,2}^{\sf S}$ is shown.

\begin{figure}[H]
\includegraphics[width=0.48\textwidth]{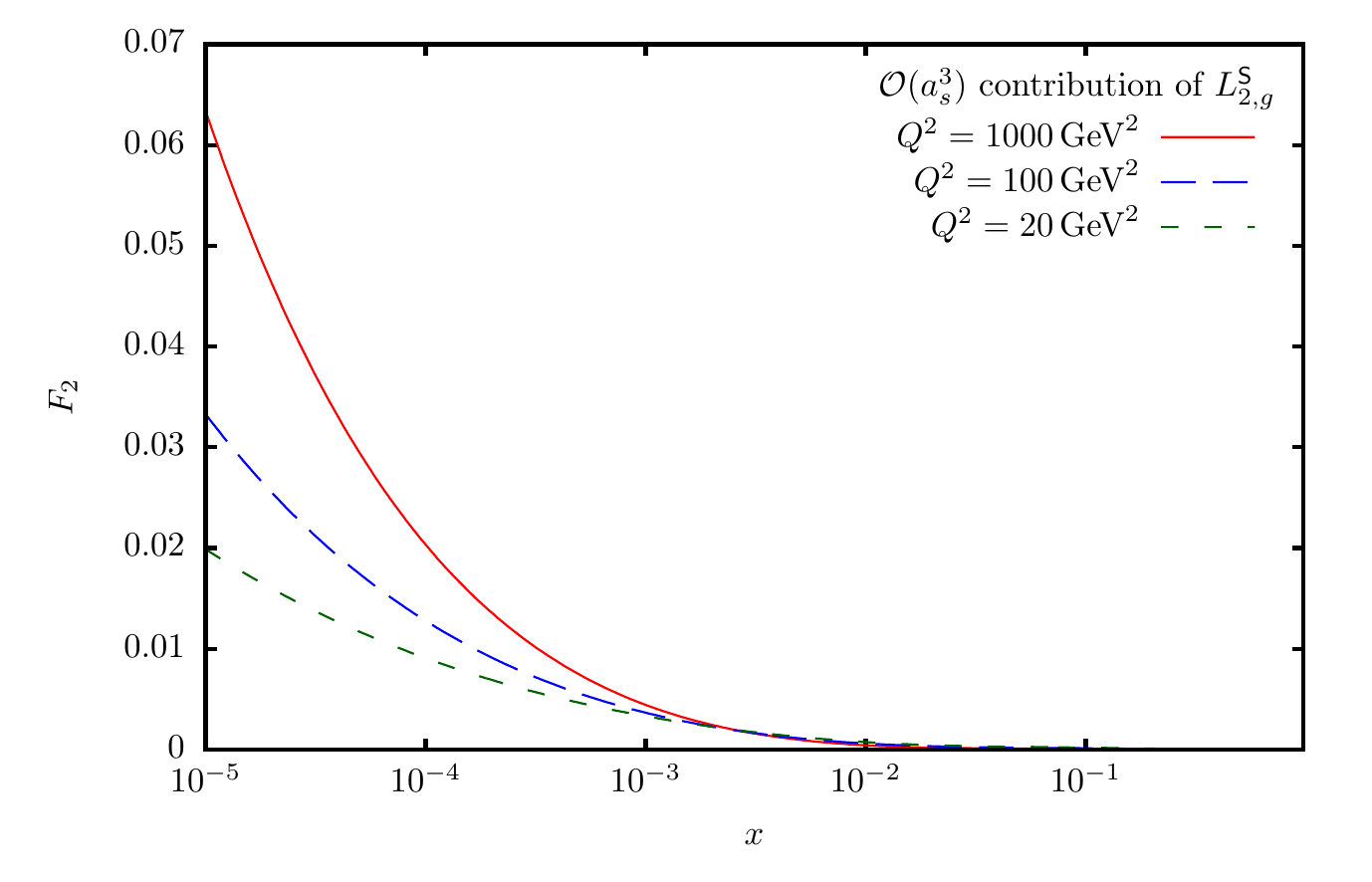}
\caption[]{The $O(a_s^3)$ contribution by $L_{g,2}^{\sf S}$ to the structure function $F_2(x,Q^2)$
using the parton distribution functions \cite{Alekhin:2013nda} and $m_c = 1.59~\GeV$; from \cite{Behring:2014eya}. 
\label{fig2}}
\end{figure}

\noindent
Comparing to the contribution of the Wilson coefficient in 3-loop order, depicted in Figure~\ref{fig2}, it turns out
that the 3-loop contribution is larger than the 2-loop contribution, which is due to a newly arising small $x$ term.

\begin{figure}[H]
\includegraphics[width=0.48\textwidth]{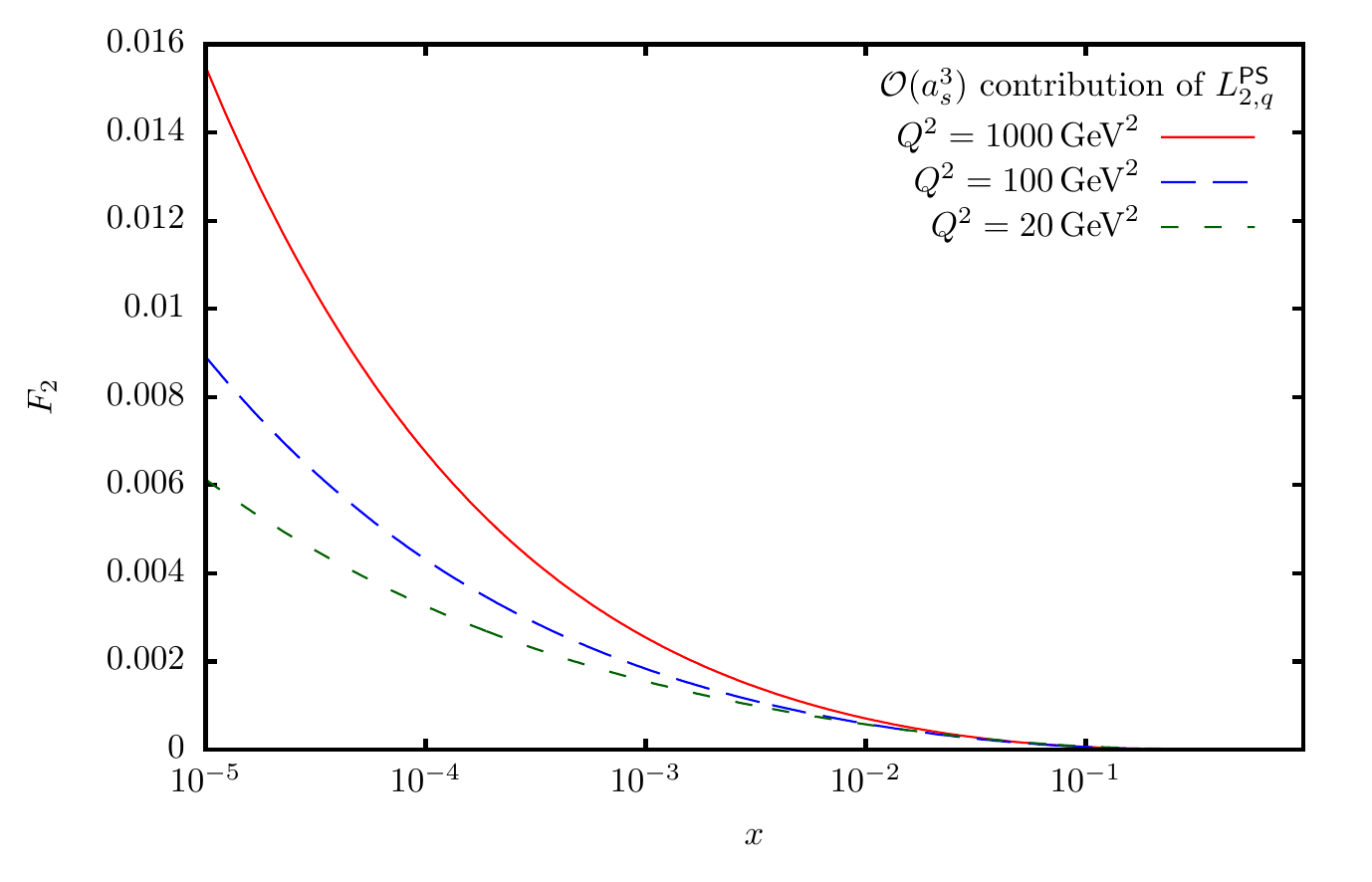}
\caption[]{The $O(a_s^3)$ contribution by $L_{q,2}^{\sf PS}$ to the structure function $F_2(x,Q^2)$ using
the parton distribution functions \cite{Alekhin:2013nda} and $m_c = 1.59~\GeV$; from \cite{Behring:2014eya}. 
\label{fig3}}
\end{figure}

\noindent
In the physical region of HERA, the widest having been explored so far, $x \geq Q^2/S , S \simeq 10^5 \GeV^2$,
the corrections are $O(0.01)$ and below. One should keep in mind that the experimental error of the $F_2$-data at HERA
is of the same size \cite{PDF}. Furthermore, measurements at the planned facilities EIC \cite{Boer:2011fh,Accardi:2012qut}
and LHeC \cite{AbelleiraFernandez:2012cc}, operating at high luminosities, will have even smaller experimental errors. 
The pure singlet contributions due to $L_{q,2}^{\rm PS}$ shown in Figure~\ref{fig3} are smaller than those of $L_{g,2}^{\rm S}$.

\begin{figure}[H]
\includegraphics[width=0.48\textwidth]{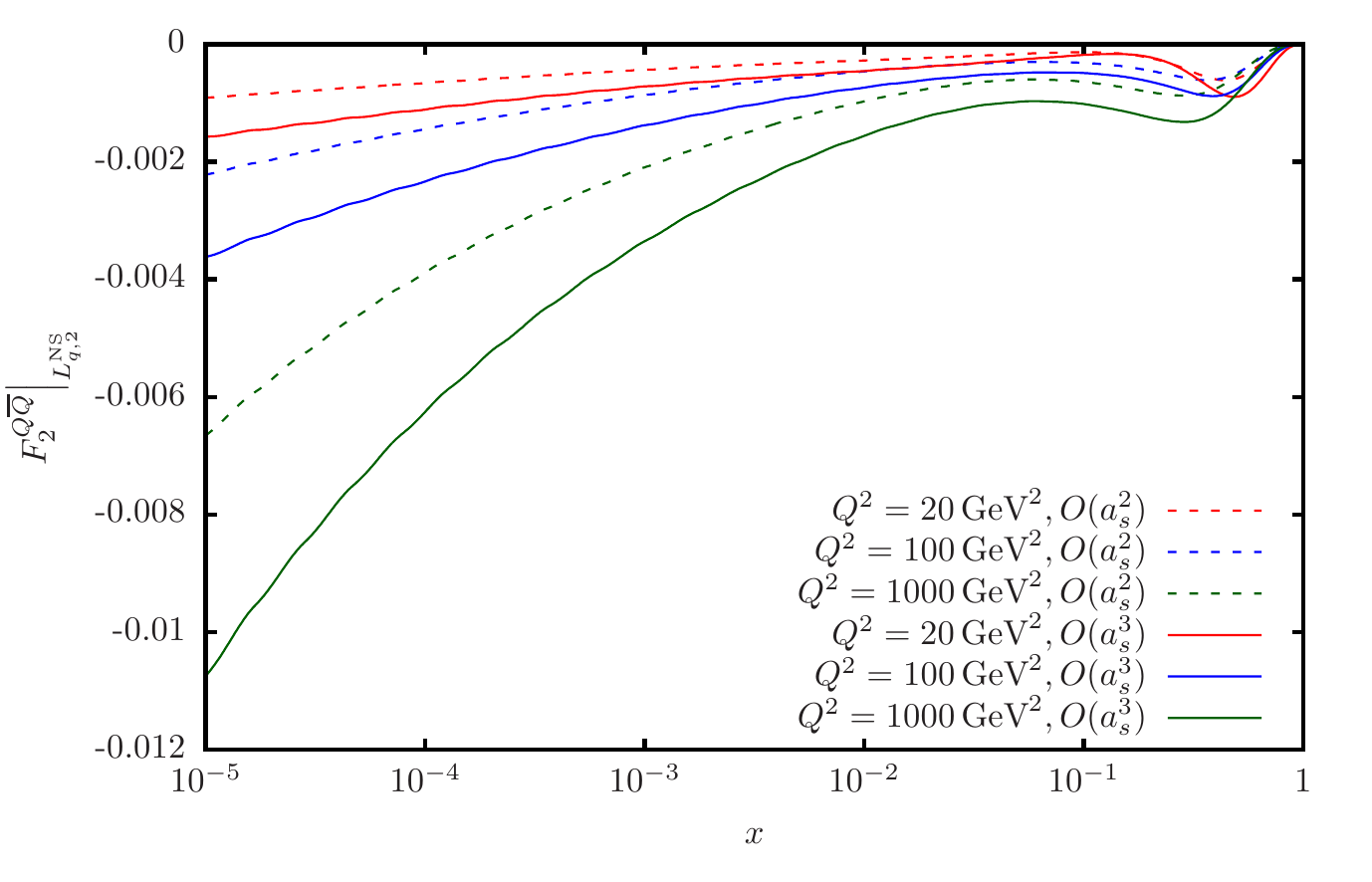}
\caption[]{The flavor non-singlet contribution of the Wilson coefficient $L_{q,2}^{\rm NS}$ to the structure function
$F_2(x,Q^2)$ at the  2- and up to the 3-loop order using the NNLO parton distribution functions of Ref.~\cite{Alekhin:2013nda}
in the on-shell scheme for $m_c = 1.59~\GeV$.  Here we do not display the $O(a_s^0)$ terms; from \cite{Ablinger:2014vwa}.}
\label{Fig:WILS1}
\end{figure}

The next terms are those due to $L_{q,2}^{\rm NS}$ depicted in Figure~\ref{Fig:WILS1}. Both the inclusive heavy flavor 2- and 
3-loop corrections are negative. The 3-loop corrections enlarge the effect towards small values of $x$. In the kinematic region at 
HERA the effects are below $\sim 0.005$.


\begin{figure}[H]
\includegraphics[width=0.48\textwidth]{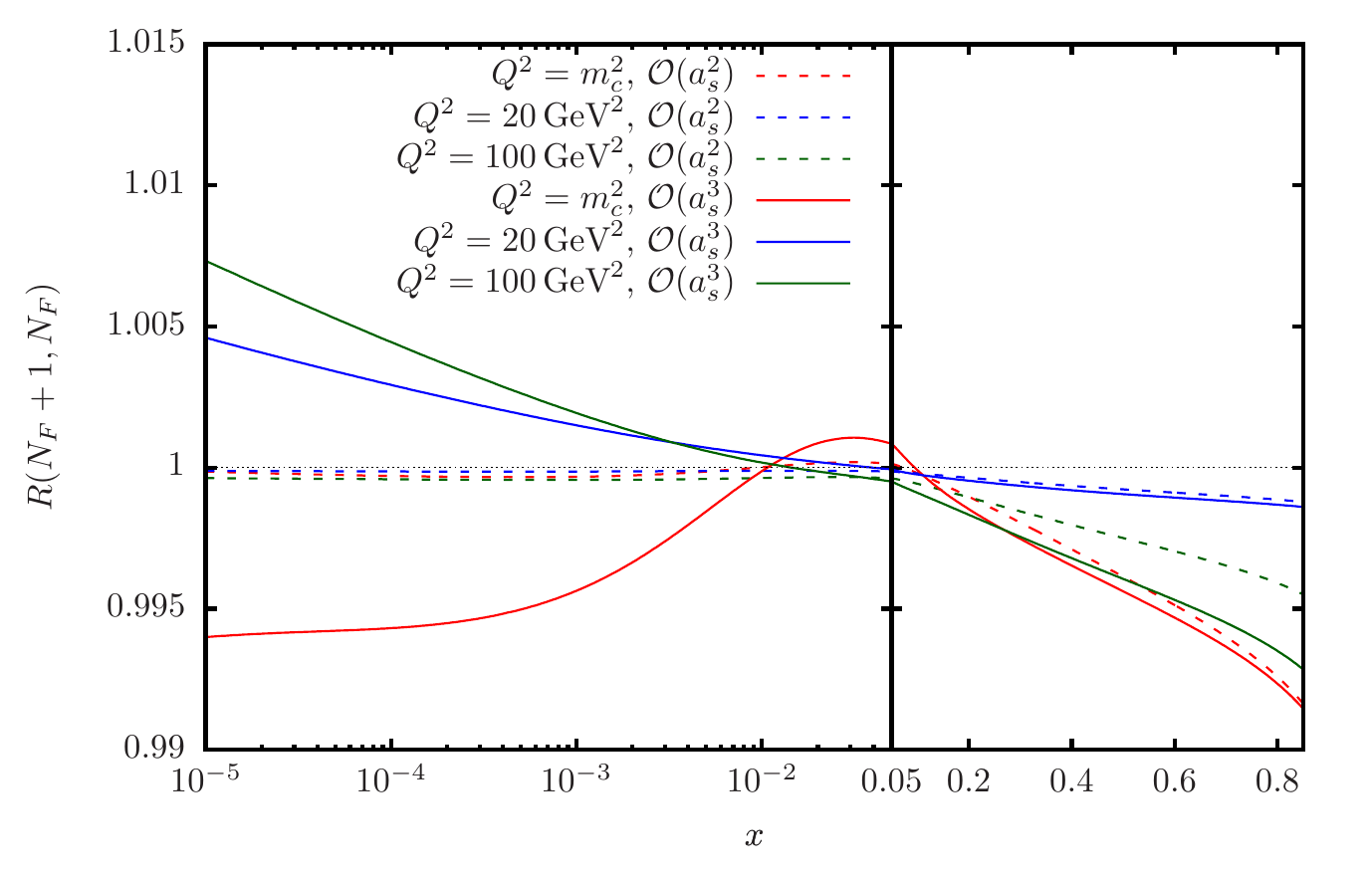}
\caption[]{
\small The ratio of the
distribution $x(u+\bar{u})$ for four and three quark flavors at 2- and 3-loop order in the variable flavor
number scheme matched at different scales of $Q^2$ as a function of $x$ using the parton distribution functions of
Ref.~\cite{Alekhin:2013nda} and the on-mass-shell
definition of the charm quark mass $m_c = 1.59~\GeV$; from \cite{Ablinger:2014vwa}.}
\label{Fig:VFNS1}
\end{figure}

\noindent
In the flavor non-singlet case all OMEs contributing to the matching relation in the VFNS up to 3-loop order (\ref{eq:VNS}) have 
been calculated. It is illustrated in Figure~\ref{Fig:VFNS1} for 2- and 3-loop order in dependence of the matching scale $Q^2$. 
While at 2-loop order the corrections are very small in the low $x$ region, it grows to large $x$ to up to O(0.005). The 3-loop 
corrections are larger due to the gluonic and singlet corrections and vary between negative and positive values of O(0.005) with 
more pronounced profiles, also in the small $x$ region. 

\begin{figure}[H]
\includegraphics[width=0.48\textwidth]{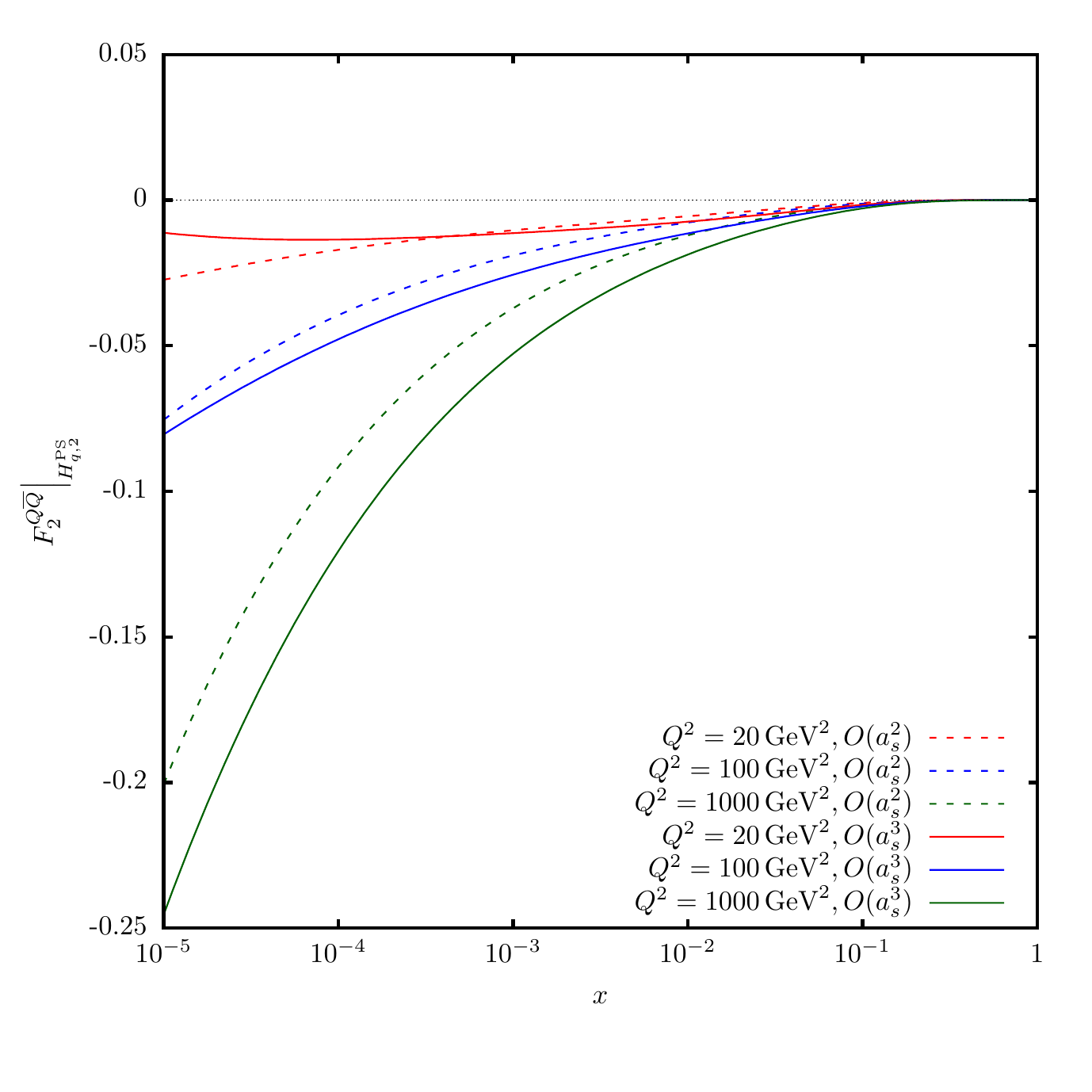}
\caption[]{The charm contribution by the Wilson coefficient $H_{Q,2}^{\rm PS}$ to the structure function
$F_2(x,Q^2)$ as a function of $x$ and $Q^2$ choosing $Q^2 = \mu^2, m_c = 1.59 \GeV$ (on-shell scheme) using the parton distribution
functions \cite{Ablinger:2014nga}.}
\label{FIG:F2c}
\end{figure}

Finally, the heavy flavor pure singlet contributions to $F_2(x,Q^2)$ are illustrated in Figure~\ref{FIG:F2c}. The corrections are 
negative both at 2- and 3-loop order and grow with $Q^2$ towards small values of $x$. In the kinematic range of HERA the 
corrections amount to $O(-0.03)$.

\vspace*{-5mm}

\subsubsection{The case of two different heavy quark masses}
\label{sec:x.1a}

\noindent
Beginning with 3-loop order Feynman integrals with two massive internal fermion lines contribute, which give rise to 
non-factorizable heavy flavor effects. In particular these diagrams cannot be attributed to a single heavy quark
species. Due to the fact that $m_c^2/m_b^2 \sim 1/10$, charm will not have decoupled to a massless fermion at the scale of
$m_b$. The VFNS generalizes in this case and one has to decouple charm and bottom at large scales together rather than 
individually \cite{BW1}. Also a series of moments for all OMEs in the 2-mass case has been calculated. The computation proceeds
along similar lines as in the single mass case. The corresponding tadpoles are calculated using the code {\tt qexp} \cite{QEXP}.
As an example we show the 6th moment of the constant part of the unrenormalized OME $A_{Qg}^{(3)}$ with $x = m_1^2/m_2^2$, 
expanding in the mass ratio $x$,

{\footnotesize
\vspace*{-7mm}
\begin{eqnarray}
\lefteqn{
\hspace*{-7mm}
a_{Qg}^{(3)}(N=6) = 
} 
\nonumber\\ && 
\hspace*{-14mm}
\frac{1}{2} \Biggl\{
        T_F^2 C_A \Bigg\{
          \frac{69882273800453}{367569090000}
          - \frac{395296}{19845} \zeta_3
          + \frac{1316809}{39690} \zeta_2
\nonumber\\ && 
\hspace*{-14mm}
          + \frac{832369820129}{14586075000} x
          + \frac{1511074426112}{624023544375} x^2
          - \frac{84840004938801319}{690973782403905000} x^3
 \N\\&&
\hspace*{-14mm}
       + \ln\Bigl( \frac{m_2^2}{\mu^2} \Bigr)    \Bigl[
         \frac{11771644229}{194481000}
          + \frac{78496}{2205} \zeta_2
          - \frac{1406143531}{69457500} x
\nonumber\\ && 
\hspace*{-14mm}
          - \frac{105157957}{180093375} x^2
          + \frac{2287164970759}{7669816654500} x^3
          \Bigr]
       + \ln^2\Bigl(\frac{m_2^2}{\mu^2}\Bigr)    \Bigl[
          \frac{2668087}{79380}
\N\\&&
\hspace*{-14mm}
          + \frac{112669}{661500} x
       - \frac{49373}{51975} x^2
          - \frac{31340489}{34054020} x^3
          \Bigr]
             + \ln^3\Bigl(\frac{m_2^2}{\mu^2}\Bigr) \frac{324148}{19845}
\N\\&&
\hspace*{-14mm}
       + \ln^2\Bigl(\frac{m_2^2}{\mu^2}\Bigr)
       \ln\Bigl(\frac{m_1^2}{\mu^2}\Bigr)   \frac{156992}{6615}
       + \ln\Bigl(\frac{m_2^2}{\mu^2}\Bigr)
       \ln\Bigl(\frac{m_1^2}{\mu^2}\Bigr)    \Bigl[
       \frac{128234}{3969}
          -  \frac{112669}{330750} x
\N\\&&
\hspace*{-14mm}
          +  \frac{98746}{51975} x^2
          +  \frac{31340489}{17027010} x^3
          \Bigr]
       + \ln\Bigl(\frac{m_2^2}{\mu^2}\Bigr)
       \ln^2\Bigl(\frac{m_1^2}{\mu^2}\Bigr) \frac{68332}{6615}
       + \ln\Bigl(\frac{m_1^2}{\mu^2}\Bigr) \Bigl[
\N\\&&
\hspace*{-14mm}
  \frac{83755534727}{583443000}
          + \frac{78496}{2205} \zeta_2
          + \frac{1406143531}{69457500} x
          + \frac{105157957}{180093375} x^2
\N\\&&
\hspace*{-14mm}
          - \frac{2287164970759}{7669816654500} x^3
          \Bigr]
       + \ln^2\Bigl(\frac{m_1^2}{\mu^2}\Bigr)  \Bigl[
           \frac{2668087}{79380}
          + \frac{112669}{661500} x
          - \frac{49373}{51975} x^2
\N\\&&
\hspace*{-14mm}
          - \frac{31340489}{34054020} x^3
          \Bigr]
       + \ln^3\Bigl(\frac{m_1^2}{\mu^2}\Bigr) \frac{412808}{19845}
\Biggr\}
+T_F^2 C_F \Bigg\{
          - \frac{3161811182177}{71471767500}
\N\\&&
\hspace*{-14mm}
          + \frac{447392}{19845} \zeta_3
          + \frac{9568018}{4862025} \zeta_2
          - \frac{64855635472}{2552563125} x
          + \frac{1048702178522}{97070329125} x^2
\N\\&&
\hspace*{-14mm}
          + \frac{1980566069882672}{2467763508585375} x^3
       + \ln\Bigl(\frac{m_2^2}{\mu^2}\Bigr)    \Bigl[
          \frac{1786067629}{204205050}
          - \frac{111848}{15435} \zeta_2
\N\\&&
\hspace*{-14mm}
          - \frac{128543024}{24310125} x
          - \frac{22957168}{3361743} x^2
          - \frac{2511536080}{2191376187} x^3
          \Bigr]
\N\\&&
\hspace*{-14mm}
              + \ln^2\Bigl(\frac{m_2^2}{\mu^2}\Bigr)    \Bigl[
          \frac{3232799}{4862025}
          + \frac{752432}{231525} x
          + \frac{177944}{40425} x^2
          + \frac{127858928}{42567525} x^3
          \Bigr]
\N\\&&
\hspace*{-14mm}
       - \ln^3\Bigl(\frac{m_2^2}{\mu^2}\Bigr)    \frac{111848}{19845}
       - \ln^2\Bigl(\frac{m_2^2}{\mu^2}\Bigr)
       \ln\Bigl(\frac{m_1^2}{\mu^2}\Bigr)      \frac{223696}{46305}
       + \ln\Bigl(\frac{m_2^2}{\mu^2}\Bigr)
       \ln\Bigl(\frac{m_1^2}{\mu^2}\Bigr)    
\N\\&&
\hspace*{-14mm}
\times \Bigl[
            \frac{22238456}{4862025}
          - \frac{1504864}{231525} x
          - \frac{355888}{40425} x^2
          - \frac{255717856}{42567525} x^3
          \Bigr]
       + \ln\Bigl(\frac{m_2^2}{\mu^2}\Bigr)
\N\\&&
\hspace*{-14mm}
       \ln^2\Bigl(\frac{m_1^2}{\mu^2}\Bigr) \frac{223696}{46305}
       + \ln\Bigl(\frac{m_1^2}{\mu^2}\Bigr)    \Bigl[
                 - \frac{24797875607}{1021025250}
          - \frac{111848}{15435} \zeta_2
\N\\&&
\hspace*{-14mm}
          + \frac{128543024}{24310125} x
          + \frac{22957168}{3361743} x^2
          + \frac{2511536080}{2191376187} x^3
          \Bigr]
\N\\&&
\hspace*{-14mm}
       + \ln^2\Bigl(\frac{m_1^2}{\mu^2}\Bigr)    \Bigl[
          \frac{3232799}{4862025}
          + \frac{752432}{231525} x
          + \frac{177944}{40425} x^2
          + \frac{127858928}{42567525} x^3
\Biggr\}
\N\\ &&
\hspace*{-14mm}
       - \ln^3\Bigl(\frac{m_1^2}{\mu^2}\Bigr) \frac{1230328}{138915}\Biggr\}
+O\left(x^4 \ln^3(x)\right).
\end{eqnarray}
}

\normalsize
\noindent
The scalar integrals contributing to all topologies of $A_{gg}^{(3)}$ in the 2-mass case at general values
of $N$ have also been calculated, which can be expressed in terms of generalized harmonic sums, where no expansion in the mass 
ratio is performed.
Indeed the expansion is possible fixing the value of $N$. Since the expansion leads to problems at general values of $N$  one
is forced to seek the complete solution.

\vspace*{-.5cm}
\section{Mathematical Aspects of Higher Loop Calculations}
\label{sec:3}

\noindent
Feynman integrals have representations in special number and function spaces, which obey a sequential order growing with the 
complexity of the integrals defined by their loop order, the number of legs and scales being involved. This has been observed in 
various calculations up to two-loop order and inclusive $2 \rightarrow 2$ processes, mainly in QCD, until the mid 1990s, cf. 
e.g. \cite{Laporta:1996mq,Hamberg:1990np}.

For zero scale quantities, like the expansion coefficients of the $\beta$-function or individual moments of operator matrix elements, one
obtains a representation over special numbers. Single scale quantities, like splitting functions or Wilson coefficients in deep-inelastic scattering or
for the Drell-Yan process lead to 1-dimensional function representations \cite{Blumlein:2005im,Blumlein:2006rr}. 

Not much is known on basis 
representations in more complex cases. The beginning of a systematic search for basis representations falls in about this time
\cite{Broadhurst:1996kc,Borwein:1999js,Vermaseren:1998uu,Blumlein:1998if} after it has been recognized that using the available function 
representations \cite{Nielsen1909,LEWIN:1958,Kolbig:1969zza,LEWIN:1981,Kolbig:1983qt,Devoto:1983tc} contained too complicated 
arguments for 
further integration or were even not complete. In the case of zero scale quantities at lower loop order in the massless case, 
multiple zeta 
values \cite{Blumlein:2009cf} are sufficient to represent the results. In massive calculations at higher loop order 
also cyclotomic zeta values \cite{Broadhurst:1998rz,Ablinger:2011te}, generalized infinite sums \cite{Ablinger:2013cf} and constants 
associated to nested (inverse) binomial sums \cite{Ablinger:2014bra,BINOM} occur. Zero scale quantities can be obtained as a fixed moment 
or in the limit $N \rightarrow \infty$ of convergent single scale quantities in Mellin space or as special values of the associated 
iterated integrals at $x = 1$ or other (algebraic) arguments.

The harmonic sums are defined by \cite{Vermaseren:1998uu,Blumlein:1998if}
{\small
\begin{eqnarray}
\label{eq:HS1}
\hspace*{-6mm}
S_{b,\vec{a}}(N) = \sum_{k=1}^N \frac{({\rm sign}(b))^k}{k^{|b|}} S_{\vec{a}}(k),~~S_\emptyset = 1,~~b, a_i \in \mathbb{Z} \backslash 
\{0\}.
\end{eqnarray}
}

\noindent
They have several generalizations. First one may define the real representations of cyclotomic sums by \cite{Ablinger:2011te}
{\small
\begin{eqnarray}
&&\hspace*{-12mm} S_{b_1,b_2,b_3;\vec{\bf{a}},}(N) = \sum_{k=1}^N \frac{({\rm sign}(b_3))^k}{(b_1 k + b_2)^{|b_3|}} 
S_{\vec{\bf{a}}}(k),~~S_\emptyset = 
1,
\\ &&
\hspace*{-12mm}
{\bf{a}} = (c_1,c_2,c_3);~b_{1}, c_1 \in  \mathbb{N} \backslash \{0\}, b_2,c_2 \in \mathbb{N}, b_{3}, c_3 \in  \mathbb{Z} \backslash 
\{0\}.
\nonumber
\end{eqnarray}
}

\noindent
The generalized (cyclotomic) sums are given by \cite{Ablinger:2013cf}\footnote{The generalized harmonic sums without cyclotomy, i.e.
for $b_2, c_2 = 0$ were given in \cite{Moch:2001zr}.}
{\small
\begin{eqnarray}
\label{eq:GHS1}
&&\hspace*{-12mm} S_{b_1,b_2,b_3;\vec{\bf{a}},}(y,\vec{x};N) = \sum_{k=1}^N \frac{y^k}{(b_1 k + b_2)^{|b_3|}} 
S_{\vec{\bf{a}}}(\vec{x};k),~~S_\emptyset = 
1,
\nonumber\\ &&
\hspace*{-12mm}
x_i,y \in \mathbb{C} \backslash \{0\},
{\bf{a}} = (c_1,c_2,c_3);~b_{1}, c_1 \in  \mathbb{N} \backslash \{0\}, b_2,c_2 \in \mathbb{N}, 
\nonumber\\ && \hspace*{-12mm}
b_{3}, c_3 \in  \mathbb{Z} \backslash 
\{0\}.
\end{eqnarray}
}

\noindent
They generalize the multiple polylogarithms studied in \cite{GON1,Borwein:1999js}. Even further generalizations contain the binomial 
coefficient $\binom{2k}{k}$ as a weight factor either in the numerator or denominator of the respective partial sum(s) 
\cite{Ablinger:2014bra}. An example is given by
{\small
\begin{eqnarray}
\sum_{i=1}^N \frac{1}{(i+1)\binom{2i}{i}}\sum_{j=1}^i \binom{2j}{j} \frac{1}{j} S_{-2}(j).
\end{eqnarray}
}

\noindent
The sums obey quasi-shuffle relations, i.e. their products are spanned by linear combinations of sums of the same class
and polynomials of sums of lower depth \cite{Reutenauer1993}. These relations depend only on the sums' index pattern, 
not on their value and are also called algebraic relations \cite{Blumlein:2003gb}. The linear combination is a sum
over all combinations of combined indices, which preserve the order of indices in the factors. For harmonic sums 
one obtains e.g.

{\small
\begin{eqnarray}
S_c S_{a,b} &=& S_{c,a,b} + S_{a,c,b} + S_{a,b,c} -  S_{c \wedge a,c} - S_{a,x \wedge b},
\\
a \wedge b &=& {\rm sign}(a) {\rm sign}(b)(|a| + |b|). \nonumber 
\end{eqnarray}
}

\noindent
Similar quasi shuffle relations are obtained by the cyclotomic \cite{Ablinger:2011te}, generalized \cite{Moch:2001zr,Ablinger:2013cf} and (inverse) 
binomial harmonic sums \cite{Ablinger:2014bra}. All these nested sums also obey structural relations, which depend on their specific structure 
beyond
their index set. They are implied by multiple integer argument and differentiation. The latter operation leads to equivalence classes of sums.
The structural relations have been worked out in Refs.~\cite{Blumlein:2009ta,Blumlein:2009fz} in the case of the harmonic sums and 
in 
\cite{Ablinger:2011te,Ablinger:2013cf} the cyclotomic and generalized (cyclotomic) sums.

One may associate iterated integrals to all these sums which are obtained by the Mellin transform
{\small
\begin{eqnarray}
\Mvec[f(x)](N) = \int_0^1 dx x^{N-1} f(x)~. 
\end{eqnarray}
}

\noindent 
In this way an alphabet of letters will be created, defining the respective Mellin transform linearly. In the case of the harmonic 
sums
the alphabet is formed by 
{\small
\begin{eqnarray}
\mathfrak{A} = \left\{\frac{1}{x}, \frac{1}{1-x}, \frac{1}{1+x}\right\}
\end{eqnarray}
}

\noindent 
yielding the harmonic polylogarithms \cite{Remiddi:1999ew}
{\small
\begin{eqnarray}
H_{b,\vec{a}}(x) = \int_0^x dx f_b(x) H_{\vec{a}}(x),~~H_\emptyset(x) = 1, f_c(x) \in \mathfrak{A}~.
\end{eqnarray}
}

\noindent 
The alphabets corresponding to the cyclotomic polylogarithms extend $\mathfrak{A}$ adding the inverse of the higher cyclotomic 
polynomials, i.e.  
{\small
\begin{eqnarray}
\hspace*{-0.7cm}
\left\{\frac{1}{1+x^2}, \frac{1}{1 - x + x^2}, \frac{1}{1+x+x^2+x^3+x^4}, ...\right\}.  
\end{eqnarray}
}

\noindent 
In addition, a series of numerator powers $x^i$ occur \cite{Ablinger:2011te}. In the case of the generalized sums the new letters
are \cite{Ablinger:2013cf}  
{\small
\begin{eqnarray}
\left\{\frac{1}{x - a_i} \right\},~~a_i \in \mathbb{R} \backslash \{0\}.
\end{eqnarray}
}

\noindent
One may even consider a combination of cyclotomic and generalized harmonic polylogarithms, as implied by the corresponding sums 
\cite{Ablinger:2013cf}. The nested (inverse) binomial sums imply square-root valued letters like 
{\small
\begin{eqnarray}
\hspace*{-0.7cm}
\left\{
\frac{1}{\sqrt{x}\sqrt{1 \pm x}},
\frac{1}{{x}\sqrt{1 \pm x}},
\frac{1}{\sqrt{2-x}\sqrt{1-x}},
...
\right\}.  
\end{eqnarray}
}

\noindent 
The special numbers associated to these classes of sums and iterated integrals are obtained in the limit $N \rightarrow \infty$ and
for $x = 1$, whenever these values exist. One formally may also associate infinite objects, cf. \cite{Blumlein:2009cf}.
The numbers obey the quasi shuffle and shuffle algebras of both spaces and a series of additional structural relations.
Counting relations of the bases for the different classes of nested sums, iterated integrals and numbers have been derived to
a large extent using Witt formulae \cite{WITT} and their corresponding modifications. The relations, basis representations, argument relations 
and conversions of the different objects quoted above are implemented in the package {\tt HarmonicSums}
\cite{Ablinger:2010kw,Ablinger:2011te,Ablinger:2013cf,Ablinger:2013hcp}.

Data analyses are often performed using Mellin-space codes, cf. e.g. \cite{Blumlein:1997em}. Here the evolution equations can be
solved analytically up to a given order in the strong coupling constant. The $N$-space representation for the structure function
can be turned into the $x$-space by a single numerical contour integral around the singularities of the problem. This representation
requests the sum representations for complex argument $N$. The analytic continuation can be derived applying the shift relations 
of the sums for $N \rightarrow N+1$ and using their asymptotic representation in analytic form \cite{Blumlein:2009ta,Ablinger:2011te,
Ablinger:2013cf,Ablinger:2014bra}. In the case of the harmonic sums also adaptive numerical representations have been derived
\cite{Blumlein:2000hw,Blumlein:2005jg,Kotikov:2005gr}. The latter may be even obtained in the case of distributions given 
numerically,
cf. e.g. \cite{Alekhin:2003ev}.

Data analyses in $x$-space require efficient numerical implementations of the iterated integrals representing the higher order 
splitting functions and Wilson coefficients. In the case of the harmonic polylogarithms the Bernoulli-improvement 
\cite{BERNOU,Gehrmann:2001pz} allows to map the expressions onto functions like $\log(1 \pm x)$ and their logarithms and 
polynomials in $x$ within the intervals $[0,\sqrt{2}-1],~[\sqrt{2}-1,1]$. The corresponding representations for more involved 
iterated integrals have still to be worked out. Numerical representations for harmonic polylogarithms were given in
\cite{Gehrmann:2001pz,Vollinga:2004sn,Maitre:2007kp,Buehler:2011ev}. Representations for generalized harmonic polylogarithms
were derived in \cite{Gehrmann:2001jv,Vollinga:2004sn}.
The mathematical structures described in this section could be systematically found calculating the corresponding Feynman integrals
in difference fields in $N$-space 
\cite{Karr:81,Schneider:01,Schneider:05a,Schneider:07d,Schneider:08c,Schneider:10a,Schneider:10b,
Schneider:10c,Schneider:13b}, 
implemented in the package {\tt Sigma} \cite{SIG1,SIG2}, or using Risch-type algorithms in $x$-space 
\cite{Risch,Ablinger:2014bra}. This also applies for new structures contained in Feynman integrals having not been revealed yet.
In this way the growing sets of function spaces describing the Feynman integrals may be uniquely found and explored. Furthermore,
for the application of the results in data analyses precise and efficient numerical implementations have to be derived.
\section[Precision Parton Distribution Functions and $\alpha_s(M_Z^2)$ and $m_c$]
{Precision Parton Distribution Functions and $\alpha_s(M_Z^2)$ and $m_c$\footnote{
Main results summarized in this section have been worked out by
S.~Alekhin, J.B., H. B\"ottcher and S.-O. Moch.
}} \label{sec:4}

\noindent
The world data on deep-inelastic scattering and related hard processes at hadron colliders like the Drell-Yan process and 
jet-production are so precise that NNLO corrections are needed to extract the parton distribution functions, the strong coupling 
constant $a_s(\mu^2) = \alpha_s(\mu^2)/(4 \pi)$ and also the charm quark mass $m_c$. In the following we give a brief survey
on the present status on the determination of these quantities.

The NNLO parton distributions have been determined by five groups (ABM, CT10, JR, MSTW, NNPDF) also resolving the flavor 
dependence of the sea-quarks. 
\begin{figure}[H]
\includegraphics[width=0.48\textwidth]{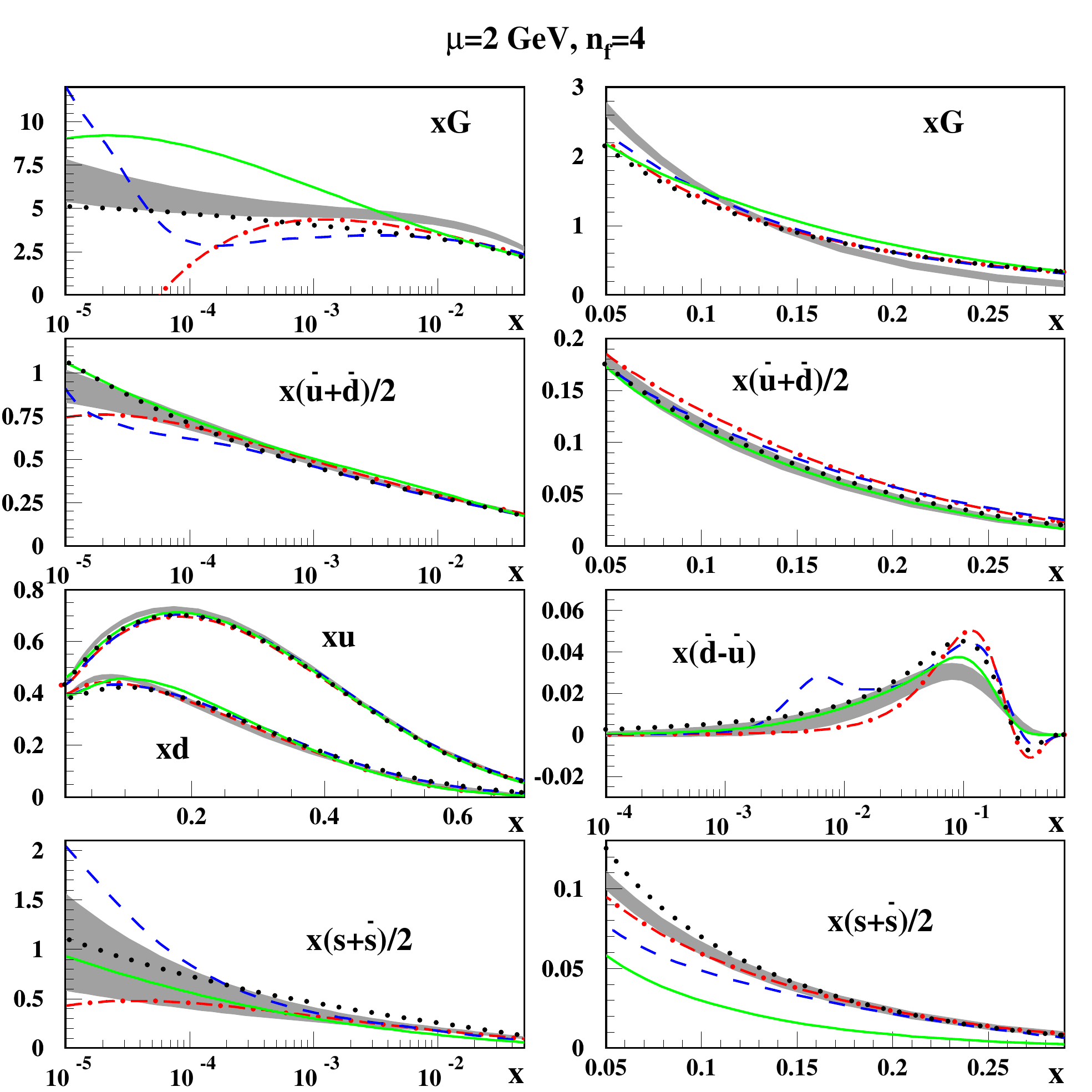}
\caption[]{
    The 1$\sigma$ band for the 4-flavor NNLO ABM12 PDFs \cite{Alekhin:2013nda} at the scale
    of $\mu=2~{\rm GeV}$ versus $x$ obtained  in this analysis (shaded area)
    compared with the ones obtained by other groups
     (solid lines: JR09~\cite{JimenezDelgado:2008hf}, dashed dots: MSTW~\cite{Martin:2009iq},
    dashes: NN23~\cite{Ball:2012cx},
    dots: CT10~\cite{Gao:2013xoa}); from Ref.~\cite{Alekhin:2013nda}.
\label{fig:pdfs1}
}
\end{figure}
\begin{figure}[H]
\includegraphics[width=0.48\textwidth]{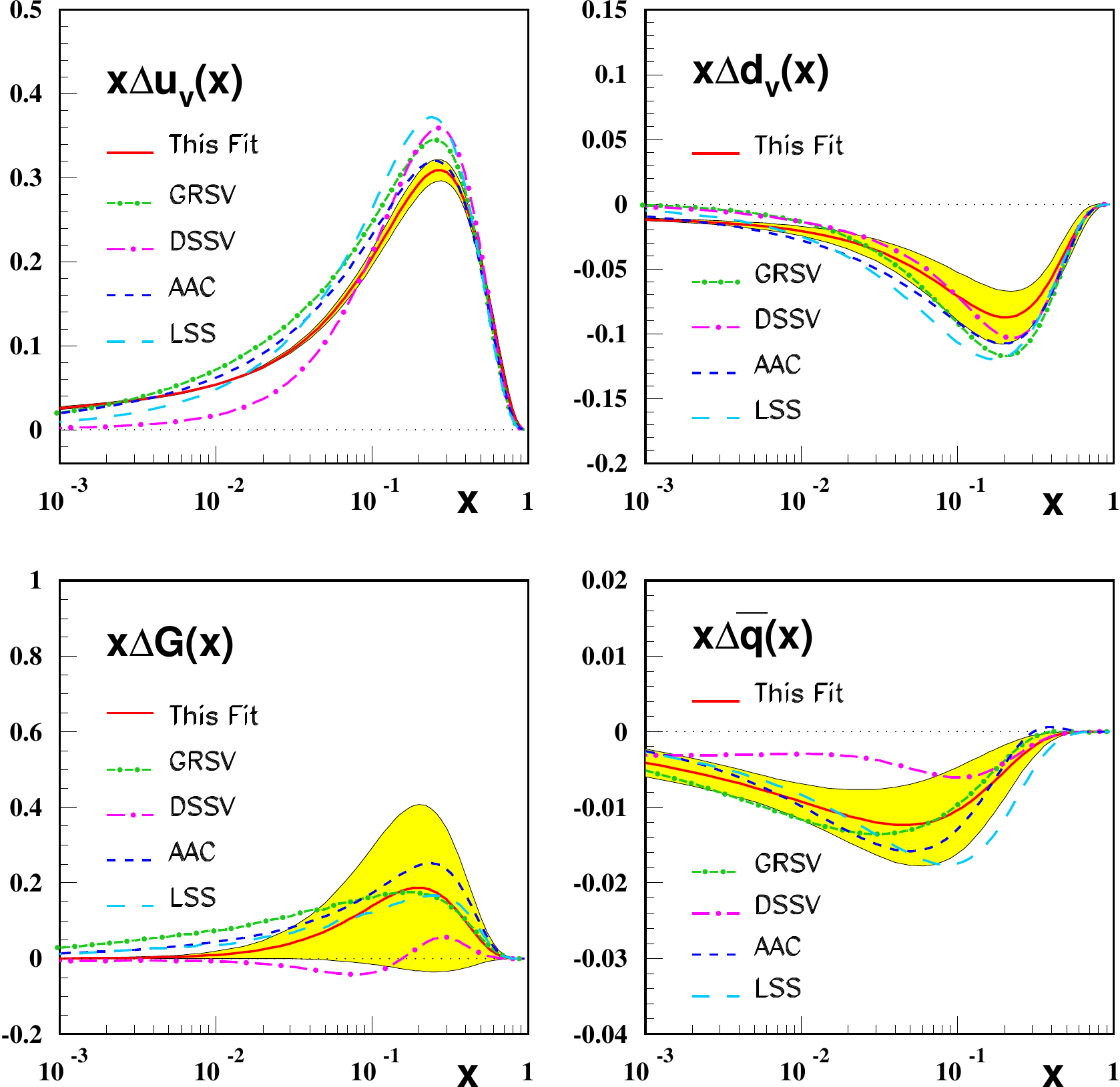}
\caption[]{\label{fig:xpdfpol}
The NLO polarized parton distributions \cite{Blumlein:2010rn} at the input scale
$\mu = 2~\GeV$ (solid line) compared to results obtained by
GRSV~(dashed--dotted line)~\cite{GRSV}, DSSV~(long dashed--dotted
line)~\cite{DSSV}, AAC~(dashed line)~\cite{AAC}, and LSS~(long dashed
line)~\cite{LSS}.
The shaded areas represent the fully correlated $1\sigma$ error bands
calculated by Gaussian error propagation; from Ref.~\cite{Blumlein:2010rn}.}
\end{figure}

\noindent
The most recent analyses were given in 
Refs.~\cite{Martin:2009iq,Ball:2012cx,Gao:2013xoa,Alekhin:2013nda,Jimenez-Delgado:2014twa}\footnote{For earlier ABM-analyses
see Refs.~\cite{Alekhin:2009ni,Alekhin:2012ig}.}. 
A recent comparison 
has been performed in \cite{Alekhin:2013nda} and the main results are shown in Figure~\ref{fig:pdfs1}. The up- and down quark 
distributions 
are well understood. In the case of the gluon distribution there are still significant differences in the small-$x$ region outside 
the present 
errors. Here future NNLO analyses of the LHC jet data based on the results of the calculation \cite{Currie:2013dwa} will lead to a decision. 
There are still also some significant differences in some analyses for the $\bar{d}-\bar{u}$ distribution, which can be well measured
in the Drell-Yan process. The strange quark distribution has the largest errors among the sea quark distributions. Future neutrino and other 
charged current data will improve it as well as better collider data. For a recent analysis see \cite{Alekhin:2014sya}.

In the polarized case so far only NLO analyses have been possible \cite{Blumlein:2010rn,GRSV,DSSV,AAC,LSS}. Very recently the NNLO 
\begin{table}[H]
{\small
\begin{center}
\begin{tabular}{|l|l|l|}
\hline
\multicolumn{1}{|c|}{ } &
\multicolumn{1}{c|}{$\alpha_s({M_Z^2})$} &
\multicolumn{1}{c|}{  } \\
\hline
Alekhin \text{\cite{Alekhin:2001ih}}  & $0.1143 \pm 0.0013$  & \\ 
BBG \text{\cite{Blumlein:2006be}}        & $0.1134 {\tiny{\begin{array}{c} +0.0019 \\ 
                                           -0.0021 \end{array}}}$ & {\rm val. analysis, N$^2$LO} 
\\
GRS \text{\cite{Gluck:2006yz}}           & $0.112 $ & {\rm val. analysis, N$^2$LO}  
\\
ABKM \text{\cite{Alekhin:2009ni}}        & $0.1135 \pm 0.0014$ & {\rm HQ:~FFNS~$N_F=3$} 
\\
JR14 \text{\cite{Jimenez-Delgado:2014twa}} & $0.1136 \pm 0.0004$ & {\rm dynamical~approach} 
\\
JR14 \text{\cite{Jimenez-Delgado:2014twa}} & $0.1162 \pm 0.0006$ & {\rm including NLO-jets}  
\\
MSTW \text{\cite{Martin:2009bu}}         & $0.1171\pm 0.0014$ &  
\\
Thorne \text{\cite{Thorne:2014toa}}      & {$0.1136$} &  DIS+DY+HT$^{*}$  
\\
ABM11$_J$ \text{\cite{Alekhin:2010iu}}   & $0.1134-0.1149$ &  Tevatron jets NLO
\\
                                         & $\pm 0.0012$    & 
\\
ABM12 \text{\cite{Alekhin:2013nda}}      & $0.1133\pm 0.0011$ & 
\\
ABM12 \text{\cite{Alekhin:2013nda}}      & $0.1132\pm 0.0011$ & (without jets)
\\
CTEQ \text{\cite{Gao:2013xoa}}           & $0.1159...0.1162$ &  \\
CTEQ \text{\cite{Gao:2013xoa}}           & {$0.1140$}& (without jets)  \\
NN21 \text{\cite{Ball:2011us}}           & $0.1174 \pm 0.0006 $ &  \\
\hline
{\rm
$e^+e^-$~{\small thr.}}~\text{\cite{Gehrmann:2012sc}}
& {{$0.1131~^{+~0.0028}_{-~0.0022}$}} & 
\\
{\rm $e^+e^-$~{\small thr.}}~\text{\cite{Abbate:2012jh}}& {{$0.1140 
\pm 0.0015$}} & 
\\
\hline
BBG \text{\cite{Blumlein:2006be}} & {{$
0.1141 {\tiny{\begin{array}{c} +0.0020 \\
-0.0022 \end{array}}}$}}
& {\rm val. analysis N$^3$LO}  
\\
\hline
World Average  & $0.1185 \pm 0.0006$ & (2013) \text{\cite{PDG13}}\\
\hline
\end{tabular}
\end{center}
\caption[]{A survey on the present status on the determination of $\alpha_s(M_Z^2)$ 
from the deep-inelastic world data and related data at NNLO and N$^3$LO.}
}
\end{table}

\normalsize
\noindent
anomalous dimensions have been calculated \cite{Moch:2014sna} and will allow for NNLO analyses in the future. In Figure~\ref{fig:xpdfpol}
we compare the present polarized parton distribution functions obtained from the inclusive and semi-inclusive world data. While a reasonable
agreement is found for the valence quark distributions the present differences are larger in the case of the gluon and sea quark 
data.
In particular the gluon distribution is strongly correlated with the value of the strong coupling constant. In this case a NNLO
analysis will lead to further improvement. In these analyses proper normalization of the denominator function of the measured asymmetries
and the treatment of the higher twist contributions is of importance, since most of the present data stem from the lower $Q^2$ domain. 

Finally, we discuss the determination of the strong coupling constant $\alpha_s(M_Z^2)$ from deep inelastic and collider data at 
NNLO.
Valence analyses have been carried out in \cite{Blumlein:2006be,Gluck:2006yz}. Here the gluon uncertainties can be avoided. Sea-quark 
tail effects were studied in \cite{Blumlein:2012se}, which yield a negligible contribution, however. An early NNLO singlet analysis was 
carried out in \cite{Alekhin:2001ih}. 

Again, the higher twist effects have to be dealt with in a careful manner. In the valence analyses
one may cut 
them away for the $\alpha_s$ determination and analyze these data separately to determine the higher twist contributions
to $F_2^{p,d}$, cf.~\cite{Blumlein:2006be,Blumlein:2008kz,Blumlein:2012se}. Today many singlet-analyses deliver comparably low values
of $\alpha_s$ \cite{Alekhin:2001ih,Alekhin:2009ni,Jimenez-Delgado:2014twa,Thorne:2014toa,Alekhin:2010iu,Alekhin:2013nda,Gao:2013xoa}, 
fully consistent with the results of the non-singlet analyses with values in the range of 0.112--0.114. Exceptions are the results
of MSTW \cite{Martin:2009bu} and NNPDF \cite{Ball:2011us}. We would like to note that also the analysis of thrust data in $e^+e^-$ annihilation
leads to small values of $\alpha_s$ \cite{Gehrmann:2012sc,Abbate:2012jh} if compared to the present world average. Theoretical uncertainties
are due to the heavy flavor treatment of $O(0.0006)$ \cite{Alekhin:2009ni} and higher order effects, comparing the central value in 
the case
of the valence N$^3$LO analysis \cite{Blumlein:2006be} with the value at NNLO of + 0.0009. The reported error of the current world average 
of $0.0006$ \cite{PDG13} is thus too small. For a survey on other determinations of $\alpha_s$ see \cite{Bethke:2011tr,Moch:2014tta,WOLF}.
Low values of $\alpha_s$ at NLO have recently been reported from ATLAS and CMS from their jet data \cite{Rabbertz:2013vxa} with
\begin{eqnarray}
\alpha_s(M_Z^2) &=& 0.111 {\tiny \begin{array}{c} + 0.0017 \\ - 0.0007 \end{array}} \\
\alpha_s(M_Z^2) &=& 0.1148 \pm 0.0055.
\end{eqnarray}
It will be interesting to see the results of the forthcoming NNLO analyses. It is needless to say that both the precise knowledge 
of $\alpha_s(M_Z)$ and the gluon distribution are instrumental for a detailed understanding of the Higgs- and top-quark production 
cross sections at the LHC \cite{Alekhin:2010dd,Alekhin:2011ey}.

A determination of the charm quark mass from deep-inelastic data has been performed in \cite{Alekhin:2012vu}, modeling the NNLO effects
\cite{Kawamura:2012cr}, assigning a corresponding theoretical error. One obtains at NNLO 
\begin{eqnarray}
\hspace*{-.7cm}
m_c(m_c) = 1.24 \pm 0.03~\text{(exp)} 
{\tiny \begin{array}{c} 
{{+ 0.03}}\\
{- 0.02} \end{array}}~\text{(scale)}
{\tiny \begin{array}{c} 
{+ 0.00}\\
{- 0.07} \end{array}}~\text{(th.)} 
\end{eqnarray}
fully compatible with the result in \cite{Chetyrkin:2009fv,KUEHN}, although with a larger error.
The theory error will be further improved after the NNLO heavy flavor corrections of Section \ref{sec:2} are fully available.

\vspace*{-4mm}
\section{Conclusions}
\label{sec:5}

\noindent
The present world data on deep-inelastic scattering and upcoming collider data from the LHC allow for a very precise determination
of the strong coupling constant $\alpha_s$ with an accuracy of 1\% accompanied with a precise determination of the parton distribution
functions. The corresponding QCD fits require NNLO analyses for which also the heavy flavor corrections have to be known at 
NNLO.
Fortunately, for the structure function $F_2(x,Q^2)$ the asymptotic 3-loop corrections can be applied at $Q^2 \gsim 23 \GeV^2$ for
the charm contributions, which yield the most important part. Recently we made important progress in calculating these corrections,
after having completed a series of moments for all quantities in 2009. All
logarithmic corrections are known and six out of eight OMEs 
are completed for general values of $N$ as well as four out of five Wilson coefficients. 

On the technical side, the calculation had to be accompanied by various innovative developments in summation and integration 
theory and new techniques in treating massive 3-loop Feynman diagrams containing local operator insertions. This also has led
to the full exploration of new mathematical function spaces. All these techniques are of importance also in other massive
calculations, e.g. for processes at the large hadron collider LHC and a forthcoming linear collider like the ILC. The ongoing 
computations are performed on a number of main frames with now typically $\sim 400$ Gbyte RAM and 20 Tbyte fast discs. With the
help of these machines, using parallel processing, we can already reduce all diagrams to master integrals. We still have to
complete the calculation of master integrals using the techniques being described in this article for the yet open OMEs 
and Wilson coefficients. Finally, also the graphs containing two different mass scales have to be calculated to complete all
the corrections.

\vspace*{2mm}
\noindent
{\bf Acknowledgment.~}~~For cooperation on the projects reported on we would like to thank 
J.~Ablinger,
S.~Alekhin,
A.~Behring,
I.~Bierenbaum,
H.~B\"ottcher,
A.~Hasselhuhn,
S.~Klein,
A.~von~Manteuffel,
S.-O.~Moch,
C.~Raab,
M.~Round,
and
F.~Wi\ss{}brock. We thank K.~Chetyrkin and
M.~Steinhauser for discussions. This work was supported in part by DFG Sonderforschungsbereich Transregio 9, Computergest\"utzte 
Theoretische 
Teilchenphysik, for which we would like to thank the Deutsche Forschungsgemeinschaft, the Austrian Science Fund (FWF) grants 
P20347-N18 and SFB F50 (F5009-N15), and the European Commission through contracts PITN-GA-2010-264564 ({LHCPhenoNet}) and 
PITN-GA-2012-316704 ({HIGGSTOOLS}).






\nocite{*}


\end{document}